\documentclass[aps,prd,preprint,12pt,superscriptaddress,nofootinbib,a4paper]{revtex4-1}

\usepackage[utf8]{inputenc}
\usepackage{amsmath,amssymb}
\usepackage[separate-uncertainty=true]{siunitx}
\usepackage{graphicx}
\usepackage[usenames,dvipsnames]{xcolor}
\usepackage[caption=false]{subfig}
\usepackage{hyperref}
\usepackage[capitalize]{cleveref}
\usepackage{booktabs}
\usepackage{tabularx}
\usepackage{xspace}
\usepackage{color}
\usepackage[normalem]{ulem}
\usepackage{slashed}
\usepackage{bbold}
\usepackage{wasysym}
\usepackage{graphicx}
\usepackage{mathrsfs}

\allowdisplaybreaks

\graphicspath{{graphics/}}

\newcommand{\eqn}{equation}
\newcommand{\lb}{\left(}
\newcommand{\rb}{\right)}
\newcommand{\be}{\beta}
\newcommand{\al}{\alpha}

\newcommand{\GeV}{{\ensuremath\rm GeV}}
\newcommand{\TeV}{{\ensuremath\rm TeV}}
\newcommand{\lam}{\lambda}
\newcommand{\MeV}{{\ensuremath\rm MeV}}
\newcommand{\pb}{{\ensuremath\rm pb}}
\newcommand{\fb}{{\ensuremath\rm fb}}

\DeclareSIUnit{\pb}{pb}
\DeclareSIUnit{\fb}{fb}
\AtBeginDocument{
\heavyrulewidth=.08em
\lightrulewidth=.05em
\cmidrulewidth=.03em
\belowrulesep=.65ex
\belowbottomsep=0pt
\aboverulesep=.4ex
\abovetopsep=0pt
\cmidrulesep=\doublerulesep
\cmidrulekern=.5em
\defaultaddspace=.5em

\newcolumntype{C}{>{\centering\arraybackslash}X}
\newcolumntype{b}{C}
\newcolumntype{s}{>{\hsize=.6\hsize}C}
\newcolumntype{R}{>{\raggedleft\arraybackslash}X}
}

\begin{document}
\bibliographystyle{hunsrt}
\date{\today}
\rightline{RBI-ThPhys-2021-25}
\title{{\Large The THDMa revisited}}
\author{Tania Robens}
\email{trobens@irb.hr}
\affiliation{Ruder Boskovic Institute, Bijenicka cesta 54, 10000 Zagreb, Croatia}

\renewcommand{\abstractname}{\texorpdfstring{\vspace{0.5cm}}{} Abstract}

\begin{abstract}
    \vspace{0.5cm}
   The THDMa is a new physics model that extends the scalar sector of the Standard Model by an additional doublet as well as a pseudoscalar singlet and allows for mixing between all possible scalar states. In the gauge-eigenbasis, the additional pseudoscalar serves as a portal to the dark sector, with a priori any dark matter spins states. The option where dark matter is fermionic is currently one of the standard benchmarks for the experimental collaborations, and several searches at the LHC constrain the corresponding parameter space. However, most current studies constrain regions in parameter space by setting all but 2 of the 12 free parameters to fixed values. \\
In this work, we perform a generic scan on this model, allowing all parameters to float. We apply all current theoretical and experimental constraints, including bounds from current searches, recent results from B-physics, in particular $B_s\,\rightarrow\,X_s\,\gamma$, as well as bounds from astroparticle physics. We identify regions in the parameter space which are still allowed after these have been applied and which might be interesting for an investigation at current and future collider machines.
\end{abstract}

\maketitle

\section{Introduction}
After the discovery of a scalar boson what complies with predictions of the Standard Model (SM) Higgs boson by the LHC experiments, the quest of additional particles that stem from possible further extensions of the SM scalar sector is an important task for the experimental collaborations. Furthermore, astroparticle observations give evidence for the existance of dark matter. In a standard cosmological description, the particle content of the SM alone does not suffice to explain these observations. Therefore, many beyond the SM (BSM) extensions in addition provide a dark matter candidate.\\
In this work, we concentrate on the model which has been proposed in \cite{Ipek:2014gua,No:2015xqa,Goncalves:2016iyg,Bauer:2017ota,Tunney:2017yfp}. It extends the scalar sector of the Standard Model by an additional doublet as well as a pseudoscalar singlet, which in turn couples to a fermionic dark matter candidate. The model therefore extends the scalar sector by 5 additional particles, which we label $H,A,a,H^\pm$ in the mass eigenbasis, as well as a fermionic dark matter candidate $\chi$. In the form discussed here, the model contains 14 free parameters after electroweak symmetry breaking, out of which 2 are fixed by the measurement of the 125 \GeV~ scalar as well as electroweak precision measurements. Furthermore, the models parameter space is subject to a large number of theoretical and experimental constraints, which we will discuss in detail below. Reference \cite{Abe:2018bpo} contains a detailed discussion as well as benchmark recommendations for the LHC experimental collaborations; see also refs \cite{Pani:2017qyd,Haisch:2018znb,Abe:2018emu,Haisch:2018hbm,Haisch:2018bby,Abe:2019wjw,Butterworth:2020vnb,Arcadi:2020gge,Argyropoulos:2021sav} for more recent work on this model. Finally, refs \cite{Aaboud:2017rzf,CMS:2018zjv,Aaboud:2018esj,Aaboud:2018fvk,Aaboud:2019yqu,CMS:2019ykj,Sirunyan:2020fwm,ATLAS:2020yzc,Aad:2020rtv,Aad:2020kep,Aad:2020aob,Aad:2021jmg,Aad:2021hjy,ATLAS:2021hbr,ATLAS:2021upq,ATLAS:2021jbf,ATLAS:2021shl,ATLAS:2021GCN,ATLAS-CONF-2021-036} report on experimental constraints from current collider searches on the parameter space in specific parameter regions\footnote{Projections for the HL-LHC sensitivity can be found in \cite{CMS-PAS-FTR-18-007,ATL-PHYS-PUB-2018-027,ATL-PHYS-PUB-2018-036,CidVidal:2018eel}.}. 
This manuscript is organized as follows. In section \ref{sec:model}, we briefly review the setup of the model, including the introduction of conventions used in this work. In section \ref{sec:const}, we discuss current theoretical and experimental bounds that need to be imposed. Section \ref{sec:scan} introduces the setup of our scan and identifies regions in parameter space that are still allowed after all constraints are taken into account. We present a first glance at possible processes at $e^+e^-$ colliders in section \ref{sec:e+e-}, and conclude in section \ref{sec:summ}. 
\section{The model}\label{sec:model}
The model discussed in this work has been introduced in references \cite{Ipek:2014gua,No:2015xqa,Goncalves:2016iyg,Bauer:2017ota,Tunney:2017yfp}, and we refer the reader to these works for a detailed discussion of the model setup. We here just list generic features for brevity. We largely follow the nomenclature of \cite{Abe:2018bpo}.

The field content of the THDMa in the gauge eigenbasis consists of two scalar fields $H_{1,2}$ which transform as doublets under the $SU(2)\,\times\,U(1)$ gauge group, and additional pseudoscalar $P$ transforming as a singlet, as well as a dark matter candidate $\chi$ which we choose to be fermionic. The two Higgs Doublet Model (THDM) part of the potential is given by
\begin{align}
V_{\text{THDM}} =& \mu_1 H_1^\dagger H_1 + \mu_2 H_2^\dagger H_2 
+ \lambda_1 (H_1^\dagger H_1)^2 + 
\lambda_2 (H_2^\dagger H_2)^2
+ \lambda_3 (H_1^\dagger H_1)(H_2^\dagger H_2)\\
&+ \lambda_4 (H_1^\dagger H_2) (H_2^\dagger H_1)
 + \left[\mu_3 H_1^\dagger H_2 + {\lambda_5} (H_1^\dagger H_2)^2 + h.c.\right]
\end{align}
Fields are decomposed according to (see also e.g. \cite{Branco:2011iw})
\begin{\eqn}\label{eq:fieldpar}
H_i\,=\,\lb \begin{array}{c} \phi_i \\ \frac{1}{\sqrt{2}} \lb v_i+\rho_i+i\,\eta_i \rb \end{array} \rb
\end{\eqn}

where
\begin{\eqn}\label{eq:vevtb}
{v_1\,=\,v\,\cos\be,\;v_2\,=\,v\,\sin\be.}
\end{\eqn}

The scalar potential is 

\begin{align}\label{eq:vp}
V_P = 
  \frac{1}{2} m_P^2 P^2 + \lambda_{P_1} H_1^\dagger H_1 P^2 +
  \lambda_{P_2} H_2^\dagger H_2 P^2 +
  (\imath b_P H_1^\dagger H_2 P + h.c.).
\end{align}
Finally, the coupling between the visible and the dark sector is mediated via the interaction
\begin{\eqn*}
\mathcal{L}_\chi\,=\,-i\,y_\chi P\bar{\chi}\gamma_5\,\chi.
\end{\eqn*}
Couplings of the scalar sector to the fermionic sector are equivalent to a THDM type II (see e.g. \cite{Branco:2011iw}) and follow from
\begin{\eqn*}
\mathcal{L}_Y\,=\,-\sum_{i=1,2}\,\left\{ \bar{Q}\,Y^i_u\,\tilde{H}_i\,u_R+\bar{Q}\,Y^i_d\,{H}_i\,d_R + \bar{L}\,Y^i_\ell\,{H}_i\,\ell_R\,+\,h.c.\right\},
\end{\eqn*}
where $Y^i_{u,d}$ denote the Yukawa matrices, $Q$ and $L$ are left-handed quark and lepton doublets, $u_R,\,d_R,\,\ell_R$ label right-handed uptype, downtype, and leptonic singlets, and $\tilde{H}\,=\,\epsilon\,H_i^*$, where $\epsilon\,=\,i\,\sigma_2$ denotes the two-dimensional antisymmetric tensor. We impose an additional $Z_2$ symmetry on the model, under which the doublets transform as $H_1\,\rightarrow\,H_1,\,H_2\,\rightarrow\,-H_2$, in order to avoid contributions from flavour changing neutral currents. We furthermore set the transformation properties of the additional fields as $P\,\rightarrow\,P,\,\chi\,\rightarrow\,-\chi$. In this work, we concentrate on the case where $Y_1^u\,=\,Y_2^d\,=\,Y_2^\ell\,=\,0$, which corresponds to a type II classification of Yukawa couplings in the THDM notation and can be achieved by requiring $u_R\,\rightarrow\,-u_R$ under the said symmetry. Note the terms $\,\sim\,\mu_3,\,b_P$ induce a soft $Z_2$ breaking.\\

After electroweak symmetry breaking, the model is characterized by in total 14 free parameters. The above potential induces standard mixing for the THDM part of the potential, which we characterize by $\cos\lb \be -\al\rb,\,\tan\be$ using standard notation. Furthermore, $V_P$ introduces a mixing between the pseudoscalar part of the THDM and the new pseudoscalar $P$, with physical states given by
\begin{\eqn*}
\binom{a}{A}\,=\,\lb \begin{array}{cc} c_\theta& -s_\theta \\ s_\theta & c_\theta  \end{array}\rb\,\binom{P}{A^0}
\end{\eqn*}
where we use the notation $c_x\,\equiv\,\cos\,x,\,s_x\,\equiv\,\sin x$ here and in the following and $A^0$ denotes the mass-eigenstate of the pseudoscalar from $V_{\text{THDM}}$ prior to mixing. We choose as free parameters
\begin{\eqn}\label{eq:pars}
v,\,m_h,\,m_A,\,m_H,\,m_{H^\pm},\,m_a,\,m_\chi,\,\cos\lb \be-\al\rb,\,\tan\be,\,\sin\theta,\,y_\chi,\,\lam_3,\,\lam_{P_1},\,\lam_{P_2}
\end{\eqn}
following the suggestions in \cite{Abe:2018bpo}. Relations between these and other quantities in the Lagrangian can be found in appendix \ref{app:invert}.\\

For the parameters in eqn (\ref{eq:pars}), $v$ and $m_h$ are fixed to be $\sim\,246\,\GeV$ and $\sim\,125\,\GeV$, respectively\footnote{In principle one can also consider the case that $m_H\,\sim\,125\,\GeV$, with $m_h$ being a lighter resonance. We discard this scenario in the following.}. We are therefore left with 12 parameters which can a priori vary, but are obviously subject to theoretical and experimental constraints. Previous studies of this model usually chose to consider scenarios where all but a few are fixed, in order to facilitate the phenomenological exploration and to provide consistent benchmark scenarios for comparison. In this work, we allow all parameters to float and determine regions in the models parameter space that are highly populated\footnote{Fine-tuned regions might exist within this model which are missed in a generic scan setup.}.

\section{Theoretical and experimental constraints}\label{sec:const}
The models parameter space is subject to a list of theoretical and experimental constraints, which we list below.
\subsection{Perturbativity, perturbative unitarity and positivity of the potential}\label{sec:thconst}
In order to guarantee positivity of the potential, several relations between the potential parameters need to be obeyed (see e.g. \cite{Abe:2018bpo,Abe:2019wjw,Arcadi:2020gge}). In the notation introduced here, they are given by
\begin{eqnarray*}
&&\lam_{1,2,P1,P2}\,\geq\,0,\,\lam_3\,\geq\,-2\,\sqrt{\lam_1\,\lam_2},\,
\lam_3+\lam_4\,-2\,\left| \lam_5\right|\,\geq\,-2\,\sqrt{\lam_1\,\lam_2}.
\end{eqnarray*}
Furthermore, we require the self-couplings in the potential to obey an upper limit, which we chose as $4\,\pi$. We therefore have
\begin{\eqn*}
|\lam_i|\,\leq\,4\,\pi,\,\left|\frac{b_P}{v}\right|\,\leq\,4\,\pi.
\end{\eqn*}
Finally, bounds from perturbative unitarity are implemented using the perturbative unitarity bounds derived in \cite{Abe:2019wjw}; we refer the reader to that work for more details as well as explicit analytic expressions. We furthermore check that eqn (3.11) of \cite{Arcadi:2020gge}
(taken from \cite{Goncalves:2016iyg}) is fulfilled. 
\subsection{Constraints from electroweak precision observables, flavour constraints, and total widths}\label{sec:ewetcconst}
These bounds are tested via an implementation in SPheno \cite{Porod:2003um,Porod:2011nf}, using an interface to Sarah \cite{Staub:2008uz,Staub:2009bi,Staub:2010jh,Staub:2012pb,Staub:2013tta}.
Constraints from measurements of electroweak precision observables are taken into account using the oblique parameters $S,T,U$ \cite{Altarelli:1990zd,Peskin:1990zt,Peskin:1991sw} as a parametrization. We take the latest fit results from the Gfitter collaboration \cite{Baak:2014ora,gfitter,Haller:2018nnx}, with central values given by
\begin{\eqn*}
S\,=\,0.04\,\pm\,0.11,\,T\,=\,0.09\,\pm\,0.14,\,U\,=\,-0.02\,\pm\,0.11
\end{\eqn*}
and the correlation matrix as given in the above references. We demand $\Delta\,\chi^2\,\leq\,8.02489$, which corresponds to a 2 $\sigma$ region around the central value determined by the SM decoupling.

Two Higgs doublet models and their extensions are also subject to strong bounds from flavour constraints, cf. e.g. \cite{Arbey:2017gmh,Haller:2018nnx}. Especially the $m_{H^\pm},\,\tan\be$ plane is strongly constrained by precision flavour observabes as e.g. $B\,\rightarrow\,X_s\,\gamma,\,B_s\,\rightarrow\,\mu^+\,\mu^-$, and $\Delta M_s$. For these variables, we consider the following allowed regions:\\
\begin{itemize}

\item{}{\bf{$B\,\rightarrow\,X_s\,\gamma$}}\\

The most recent results on the theoretical calculation of this variable, including higher order calculations up to NNLO in QCD, have been presented in \cite{Misiak:2020vlo}, superseding previous results \cite{Misiak:2015xwa}. In particular, the charged scalar mass is pushed to even higher masses. We implement the lower bound in this variable using a two-dimensional fit function \cite{mm} that reflects the bounds derived in \cite{Misiak:2020vlo}. \\

\item{}{\bf $B_s\,\rightarrow\,\mu^+\,\mu^-$}\\
The ATLAS, CMS and LHCb combination using 2011-2016 LHC data for the branching ratio  $B_s\,\rightarrow\,\mu^+\,\mu^-$ is given by \cite{ATLAS-CONF-2020-049}
\begin{\eqn}\label{eq:brsmumuexp}
\lb B_s\,\rightarrow\,\mu^+\mu^-\rb^\text{comb}\,=\,\left[2.69^{+0.37}_{-0.35}\right]\times\,10^{-9}.
\end{\eqn}

In principle, theoretical predictions depend on all novel scalar masses (see e.g. \cite{Cheng:2015yfu}). 

In the decoupling limit, Spheno renders the SM output
\begin{\eqn*}
B_s\,\rightarrow\,\mu^+\mu^-\,=\,2.96\,\times\,10^{-9},  
\end{\eqn*}
while the most recent theoretical calculation yields \cite{Beneke:2019slt}
\begin{\eqn*}
B_s\,\rightarrow\,\mu^+\mu^-\,=\,\lb 3.66\,\,\pm\,0.14 \rb\times\,10^{-9}.
\end{\eqn*}
Note that this disagrees with the experimental result (eqn. (\ref{eq:brsmumuexp})) by $2.5\,\sigma$.

In order to account for the missing higher order corrections, we apply a multiplicative correction factor to the Spheno result.

Assuming again a similar error in the theory calculation, and taking into account the additional higher order shift to the Spheno result, we then have
\begin{\eqn*}
\lb B_s \rightarrow\,\mu^+\mu^- \rb^{\text{Spheno}}\,\in\left[ 1.26;3.14 \right]\,\times\,10^{-9},
\end{\eqn*}
where we allow for a $3\sigma$ deviation.

\item{}{\bf $\Delta M_s$}\\
The most recent experimental result for this value is given by $\Delta M_s\,(\text{ps}^{-1})\,=\,17.757\,\pm\,0.020\,\pm\,0.007$ \cite{Amhis:2019ckw}.

We use the following derivation for the SM prediction for this value:
We start with eqn. (31) of \cite{Lenz:2010gu}
\begin{\eqn*}
\Delta M_s\,=\,17.24 \text{ps}^{-1}\,\lb\frac{\left|V_{tb}\,V_{ts}^*\right|}{0.04}\rb^2\,\frac{S\lb \overline{m}_t^2/m_W^2\rb}{2.35}\,\frac{f^2_{B_s}\,\mathcal{B}_{B_s}}{\lb 0.21\GeV\rb^2}\,\left| \Delta_s\right|
\end{\eqn*}
which shows the dependency of this variable on external, experimentally determined parameters. The above expression was updated to \cite{Lenz:2011ti,ulidisc}
\begin{\eqn*}
\Delta M_s\,=\,17.3 \text{ps}^{-1}\,\frac{f^2_{B_s}\,\mathcal{B}_{B_s}}{\lb 211.84 \MeV\rb^2}\,\frac{\left| V_{cb}\right|^2}{0.04089^2}.
\end{\eqn*} 
For input values \cite{Aoki:2019cca,Dowdall:2019bea,Zyla:2020zbs,Aoki:2021kgd}
\begin{\eqn*}
f_{B_s}\,=\,\lb 230.3\,\pm\,1.3\rb \MeV,\,\widehat{\mathcal{B}}_{B_s}\,=\,1.232\,\pm\,0.053 ,\,V_{cb}\,=\,\lb 42.2 \,\pm\,0.8 \rb\,\times\,10^{-3},
\end{\eqn*}
together with the conversion factor $\widehat{\mathcal{B}}_{B_s}\,=\,1.52319\,\mathcal{B}_{B_s}$ \cite{Lenz:2011ti}, we then obtain

\begin{\eqn*}
\Delta M_s\,=\,\lb 17.61\,\pm\,1.05\rb\, \text{ps}^{-1},
\end{\eqn*}
where we furthermore assumed an intrinsic error of 0.2 on the central value \cite{ulidisc}, which mainly stems from uncertainties of the top mass \footnote{Neglecting this leads to a slightly smaller error of $\pm\,1.03\,\text{ps}^{-1}$.}. 
The Spheno decoupling result is given by $\lb\Delta M_s\rb^{\text{Spheno}}\,=\,18.801 \,\text{ps}^{-1}$.

Assuming the new physics contributions to be multiplicative, while keeping the error additive\footnote{TR wants to thank M. Mikolaj for useful discussions regarding this point.}, we then have

\begin{\eqn}\label{eq:dmbs_uli}
\left[\Delta M_s\,(\text{ps}^{-1})\right]^{\text{Spheno}}\,\in\left[ 16.71; 21.19 \right]\,\text{ps}^{-1}
\end{\eqn}
as an allowed $2\,\sigma$ range for this value.
\end{itemize}

Finally, although not strictly required, we impose an upper bound on the width over mass ratio on all new scalars. For the 125 \GeV~ candidate, we require that \cite{Sirunyan:2019twz}
\begin{\eqn}\label{eq:125w}
\Gamma_{h,125}\,\leq\,9\,\MeV.
\end{\eqn}
For all other scalars, we impose
\begin{\eqn}\label{eq:widthrel}
\Gamma/ M\,\leq\,0.5.
\end{\eqn}
We consider this a very lenient bound in order to allow a detailed investigation of the parameter space; in practise, ratios larger than $10-20\%$ already strongly question the applicability of perturbative treatment for such states.
\subsection{Collider bounds}\label{sec:collconst}

Agreement with current measurements of the 125 \GeV~ signal strength as well as agreement with null-results from a large number of searches have been tested by interfacing the model with HiggsBounds \cite{Bechtle:2008jh,Bechtle:2011sb,Bechtle:2013gu,Bechtle:2013wla,Bechtle:2015pma,hb,Bechtle:2020pkv} and HiggsSignals \cite{Stal:2013hwa,Bechtle:2013xfa,Bechtle:2014ewa,hb,Bechtle:2020uwn}. Furthermore, the model is additionally constrained by dedicated searches by the LHC experiments; we have implemented the corresponding bounds in a pseudo-approximate approach and leave a detailed recast analysis for future work. All production cross sections have been produced using  Madgraph5 \cite{Alwall:2011uj}, with the UFO model provided in \cite{Bauer:2017ota,ufo}.
\begin{itemize}
\item{}{\bf Bounds from $\ell^+\ell^-\,+\,\text{MET}$ searches}\\

We consider the experimental bounds presented in \cite{Sirunyan:2020fwm}, which corresponds to a CMS search for this channel mediated via $Za$ production using full Run 2 luminosity. 
For the parameters specified in figure 8 in that work, we generate parton-level events. In order to marginally mimick the experimental cuts, we have imposed
\begin{eqnarray*}
p_{\perp,\ell}\,>\,25\,\GeV,&\,\slashed{E}_\perp> 80\,\GeV,&m_{\ell\ell}\,\in\,\left[76,\,106\right] \GeV,\\ p_\perp^{\ell\ell}\,>\,60\,\GeV,&|\eta_\ell|\,\leq\,2.4,&\Delta R_{\ell\ell}\,\leq\,1.8
\end{eqnarray*}
at parton-level. Results for two specific parameter points using the above cuts are given in table \ref{tab:refvalsllmet}.
\begin{center}
\begin{table}[tb!]
\begin{center}
\begin{tabular}{c|c|c|c||c}
$m_a$&$m_A$&$\sin\theta$&$\tan\be$&$\sigma[\pb]$\\ \hline
220 \GeV& 400 \GeV&0.35&1&0.01066(3)\\
320 \GeV&650\GeV&0.35&1&0.00332(3)\\
420 \GeV&1000\GeV&0.35&1&0.0007296(9)
\end{tabular}
\caption{\label{tab:refvalsllmet} Results for parton-level generation with rudimentary cuts for points which are excluded by $\ell^+\ell^-+\slashed{E}_\perp$ searches in \cite{Sirunyan:2020fwm}. Other values are set to $m_H\,=\,m_{H^\pm}\,=\,m_A,\,m_\chi\,=\,10\,\GeV,\,\cos(\be-\al)\,=\,0,\,\lam_3\,=\,\lam_{P_1}\,=\,\lam_{P_2}\,=\,3,\,y_\chi=1.$}
\end{center}
\end{table}
\end{center}
A first estimate for sensitivity regions would therefore be to only allow parameter points with 
\begin{\eqn}\label{eq:llpart}
\sigma\,\leq\,0.0007314\,\pb,
\end{\eqn} 
and simulating all parameter points using the cuts specified above. However, this is especially computationally intensive. Therefore, it is useful to consider the main channel leading to the above signature, which is
\begin{\eqn*}
p\,p\,\rightarrow\,H\,\rightarrow\,a\,Z
\end{\eqn*}
with subsequent decays $a\,\rightarrow\,\chi\,\bar{\chi},\,Z\,\rightarrow\,\ell^+\,\ell^-$. The value from factorized onshell production times decay branching ratios into the $Z\,\chi\,\bar{\chi}$ final state for the above benchmark points is given in table \ref{tab:llmetdet}.
\begin{center}
\begin{table}[tb!]
\begin{center}
\begin{tabular}{c|c|c|c|c|c||c}
$m_a$&$m_A$&$m_H$&$\sigma_H\,[\pb]$&$H\,\rightarrow\,a\,Z$&$a\,\rightarrow\,\chi\bar{\chi}$&$\sigma_{Z\chi\bar{\chi}}^\text{fac}[\pb]$\\ \hline
220 \GeV&400 \GeV& 400 \GeV&3.70(1)&0.11&1&0.407(1)\\
320 \GeV&650 \GeV& 650 \GeV&0.534(1)&0.13&1&0.0694(1)\\
420 \GeV&1000 \GeV& 1000 \GeV&0.0497(2)&0.28&0.86&0.01197(5)
\end{tabular}
\caption{\label{tab:llmetdet} Parameter points in table \ref{tab:refvalsllmet}, relevant quantities. Last column is derived from previous entries.}
\end{center}
\end{table}
\end{center}
We therefore could use as a naive condition that
\begin{\eqn}\label{eq:llcond1}
\sigma^\text{fac}_{Z\chi\,\overline{\chi}}\,\equiv\,\sigma_{pp\,\rightarrow\,H}\,\times\,\text{BR}_{H\,\rightarrow\,Z\,a}\,\times\,\text{BR}_{a\,\rightarrow\,\chi\,\bar{\chi}}\,\leq\,0.01207.
\end{\eqn}
A more thorough inspection shows that the above bound has to be modified to
\begin{\eqn*}
\sigma^\text{fac}_{Z\chi\,\overline{\chi}}\,\leq\,0.009085\,\pb
\end{\eqn*}
in order to capture all points from our scan which obey eqn (\ref{eq:llpart}).

\item{}{\bf Bounds from $h\,+\,\text{MET}$ searches}\\
The above channel for this model has been investigated in \cite{ATLAS:2021jbf,ATLAS:2021shl} by the ATLAS collaboration using the full Run 2 dataset. We found that results for the $b\,\bar{b}$ decay of the $h$ \cite{ATLAS:2021shl} supersede those of the $\gamma\gamma$ decay \cite{ATLAS:2021jbf}, and therefore concentrate on the former in the following. Note that the search has been separated into a resolved and an unresolved region; for the resolved, we require $\slashed{E}_\perp\,>\,500\,\GeV$, while for the resolved we have $\slashed{E}_\perp\,\in\,\left[150;\,500\right]\,\GeV,\,p_{\perp,h}\,>\,100\,\GeV$. We show the values using these cuts for some sample points in table \ref{tab:hbbnew}.

\begin{center}
\begin{table}[tb!]
\begin{center}
\begin{tabular}{c|c||c|c }
$m_a$&$m_A$&$\sigma_\text{res}[\pb]$&$\sigma_\text{unres}[\pb]$\\ \hline
640 \GeV&1350 \GeV&0.001898 (2)&0.0003709 (5)\\
300 \GeV& 500 \GeV& 0.00493 (2)&$6.79 (1)\,\times\,10^{-7}$\\
180 \GeV& 1800 \GeV&0.00994 (8)&0.0005817 (2)
\end{tabular}
\caption{\label{tab:hbbnew} Results for parton-level generation with rudimentary cuts for points which are excluded by $h+\slashed{E}_\perp$ searches in \cite{ATLAS:2021shl}. Other values are set to $m_H\,=\,m_{H^\pm}\,=\,m_A,\,m_\chi\,=\,10\,\GeV,\,\cos(\be-\al)\,=\,0,\,\lam_3\,=\,\lam_{P_1}\,=\,\lam_{P_2}\,=\,3,\,y_\chi=1,\,\sin\theta\,=\,0.35,\,\tan\be\,=\,1$. $\sigma_\text{res/ unres}$ refer to the resolved/ unresolved region as discussed in the text.}
\end{center}
\end{table}
\end{center}
\begin{center}
\begin{table}[tb!]
\begin{center}
\begin{tabular}{c|c||c|c }
$m_a$&$m_A$&$\sigma_\text{res}[\pb]$&$\sigma_\text{unres}[\pb]$\\ \hline
500 \GeV&1200 \GeV&0.00466 (1)&0.000357 (1)\\
180 \GeV& 1550 \GeV& 0.002816 (1)&0.001366 (2)\\
300 \GeV& 500 \GeV&0.00493 (2)&$6.79 (1)\, \times\,10^{-7}$
\end{tabular}
\caption{\label{tab:hbbnewnew} Results for parton-level generation with rudimentary cuts for points which are excluded by $h+\slashed{E}_\perp$ searches in \cite{ATLAS:2021shl}. Other values are set to $m_H\,=\,m_{H^\pm}\,=\,m_A,\,m_\chi\,=\,10\,\GeV,\,\cos(\be-\al)\,=\,0,\,\lam_3\,=\,\lam_{P_1}\,=\,\lam_{P_2}\,=\,3,\,y_\chi=1,\,\sin\theta\,=\,0.35,\,\tan\be\,=\,1$. $\sigma_\text{res/ unres}$ refer to the resolved/ unresolved region as discussed in the text.}
\end{center}
\end{table}
\end{center}
A priori, the dominant production times decay stems from 
\begin{\eqn}\label{eq:hmetviaA}
p\,p\,\rightarrow\,A\,\rightarrow\,h\,a\,\rightarrow\,h\,\chi\,\bar{\chi},
\end{\eqn}
where additional contributions can come from
\begin{\eqn}\label{eq:hmetviaAa}
p\,p\,\rightarrow\,A (a)\,\rightarrow\,h\,A (a)\,\rightarrow\,h\,\chi\,\bar{\chi},
\end{\eqn}
i.e., non-resonant production of $A (a)$ and subsequent decays.

From table \ref{tab:hbbnewnew}, we see that we can roughly exclude points for which the conditions
\begin{\eqn}\label{eq:hmetbounds}
\sigma_{h+\slashed{E}_\perp,\text{res}}\,\leq\,0.002818\,\pb,\;\sigma_{h+\slashed{E}_\perp,\text{unres}}\,\leq\,6.81\,\times\,10^{-7}\,\pb
\end{\eqn}
are not fulfilled\footnote{We again want to emphasize that this only corresponds to a very rough sensitivity estimate. In particular, different regions of the parameter space can be more sensitive to the resolved/ unresolved region.}.

For the two points giving the bounds in eqn (\ref{eq:hmetbounds}), we have in a factorized approach
\begin{eqnarray*}
\sigma_{180,1550}^\text{fac}&=&\sigma_A\,\times\,\text{BR}_{A\,\rightarrow\,h\,a}\,\times\,\text{BR}_{a\,\rightarrow\,\chi\,\bar{\chi}}\,=\,0.001728 (2)\,\pb,\\
\sigma_{300,500}^\text{fac}&=&\sigma_A\,\times\,\text{BR}_{A\,\rightarrow\,h\,a}\,\times\,\text{BR}_{a\,\rightarrow\,\chi\,\bar{\chi}}\,=\,0.1158 (1)\,\pb.
\end{eqnarray*}
A more detailed investigation show that for the first parameter point, (\ref{eq:hmetviaA}) provides around $62\%$ of the cross-section in the resolved region, while for the second, the processes in (\ref{eq:hmetviaAa}) mainly contribute. In the following, we require that

\begin{\eqn*}
\sigma_\text{fac}\,=\,\sigma_A\,\times\,\text{BR}_{A\,\rightarrow\,a\,h}\,\times\,\text{BR}_{a\,\chi\,\bar{\chi}}\,\leq\,0.001732\,\pb.
\end{\eqn*}
For our sample points, we have explicitely checked that this value cuts out all parts of parameter space where $\sigma_{h+\slashed{E}_\perp}\,\geq\,0.002818\,\pb$.

\item{}{\bf Bounds from $H^+\,\bar{t}b,\,H^+\,\rightarrow\,t\,\bar{b}$ searches}\\
Searches for the THDM sector of the model can be constrained by the above process, which has been presented e.g. by the ATLAS collaboration \cite{Aad:2020kep} using an integrated luminosity of $139\,\fb^{-1}$, which we compare to the upper bound on the cross section as available from Hepdata \cite{hephptb}. Even a comparison with the production cross section for the process
\begin{\eqn*}
p\,p\,\rightarrow\,H^+\,\bar{t}b
\end{\eqn*}
prior to the charged scalars decay shows that this search is not yet sensitive in the parameter space investigated here. This is mainly caused by the lower limit on $\tan\be$ from $B\,\rightarrow\,X_s\,\gamma$ which preselects point with higher $\tan\be$ values.
\item{}{\bf Bounds from $Wt+\,\text{MET}$ searches}\\
The ATLAS collaboration has presented results for this model in the above channel using full Run2 data \cite{ATLAS:2020yzc}. They give results for bounds derived from single and double top production, divided into several search regions. We here implemented their results into grid-like upper bounds on the total production cross section from single-top like processes, with values taken from figure 28 of the auxiliary material available from \cite{atlaux}. In the mass region considered here, the smallest upper bound is around $32\,\fb$\footnote{Note that including $t\,\bar{t}$ results will improve these bounds. However, here theoretical cross sections which were e.g. used in figure 31 of the above auxiliary material includes detector simulation and cuts; we thank the authors of the above reference for useful discussions regarding this point. Including these would require a recast-type investigation, which is beyond the scope of the current work.}.

\item{\bf Bounds from $t\,\bar{t}/ b \bar{b}+\,\text{MET}$ final states}\\
The $t\bar{t}$ final states have been investigated in \cite{Aad:2020aob,Aad:2021hjy} using full Run 2 data. We note that the simulation for the signal a different model file was used, and cross sections were rescaled to NLO predictions \cite{Backovic:2015soa}. In the comparison here, we stick to leading order, as QCD higher order corrections are universal for both models, differing only in the electroweak sector, as long a dominant production is assumed via an s-channel mediator. In table \ref{tab:tt2}, we give our predictions for the theory cross section predictions shown in figure 15 of that reference. Note that for generation we used $\sin\theta\,=\,1/\sqrt{2}$, as otherwise no significant coupling to both the SM and the dark sector can be achieved. The values in table \ref{tab:tt2} have been rescaled accordingly\footnote{The coupling of $a$ to $t\,\bar{t}$ is rescaled by $\sin\theta$, while the coupling to $\chi\,\bar{\chi}$ is rescaled by $\cos\theta$. We therefore, in comparison to the calculating in \cite{Aad:2020aob} for $g=1$ obtain an additional factor $\frac{1}{4}$ in the results.}. We use $\slashed{E}_\perp\,>\,230\,\GeV$.
\begin{center}
\begin{table}[tb!]
\begin{center}
\begin{tabular}{c||c||c|c}
$m_a\,[\GeV]$&$\sigma[\pb]$&fac&$\sigma^\text{excl}[\pb]$\\ \hline
60&0.0570 (2)&0.45&0.02565 (9)\\
100&0.0521 (3)&0.5&0.0261 (2)\\
200&0.0371 (2)&0.8&0.0297 (2)\\
300&0.0241 (1)&1.02&0.0246 (1)\\
400&0.00868 (4)&4&0.0347 (2)
\end{tabular}
\caption{\label{tab:tt2} Results for parton-level generation with rudimentary cuts for points which are excluded by $t\bar{t}+\slashed{E}_\perp$ searches in \cite{Aad:2020aob} via $a$ mediation. Other values are set to $m_H\,=\,m_A\,=\,m_{H^\pm}\,=\,600\,\GeV,\,m_\chi\,=\,1\,\GeV,\,\cos(\be-\al)\,=\,0,\,\lam_3\,=\,\lam_{P_1}\,=\,\lam_{P_2}\,=\,3,\,y_\chi=1,\,\tan\be\,=\,1.$ The factor corresponds to a rough estimate of the multiplicative factor from figure 15 of that reference. The last column gives the actual exclusion cross section.}
\end{center}
\end{table}
\end{center}

Following the same logic as before, we see that we can naively exclude points where 
\begin{\eqn*}
t\bar{t}+\slashed{E}_\perp:\,\sigma\,\leq\,0.0248\,\pb
\end{\eqn*}

For the search in \cite{Aad:2021hjy}, we use $\slashed{E}_\perp\,\gtrsim\,110\,\GeV$. Cross-sections were calculated as above, and the respective values are given in table \ref{tab:tt3}.
\begin{center}
\begin{table}[tb!]
\begin{center}
\begin{tabular}{c||c||c|c}
$m_a\,[\GeV]$&$\sigma[\pb]$&fac&$\sigma^\text{excl}[\pb]$\\ \hline
60&0.2028 (8)&0.24&0.0487 (2) \\
100&0.1716 (8)&0.25&0.0429 (2)\\
200&0.1016 (4)&0.5&0.0508 (2)\\
300&0.0576 (3)&0.8&0.0461 (2)\\
400&0.01900 (8)&1.4&0.0266 (1)
\end{tabular}
\caption{\label{tab:tt3} Results for parton-level generation with rudimentary cuts for points which are excluded by $t\bar{t}+\slashed{E}_\perp$ searches in \cite{Aad:2021hjy} via $a$ mediation. Other values are set to $m_H\,=\,m_A\,=\,m_{H^\pm}\,=\,600\,\GeV,\,m_\chi\,=\,1\,\GeV,\,\cos(\be-\al)\,=\,0,\,\lam_3\,=\,\lam_{P_1}\,=\,\lam_{P_2}\,=\,3,\,y_\chi=1,\,\tan\be\,=\,1.$ The factor corresponds to a rough estimate of the multiplicative factor from figure 15 of that reference. The last column gives the actual exclusion cross section.}
\end{center}
\end{table}
\end{center}
This leads to
\begin{\eqn*}
t\bar{t}+\slashed{E}_\perp:\,\sigma\,\leq\,0.0268\,\pb
\end{\eqn*}
for this $\slashed{E}_\perp$ cut.

The $b\bar{b}$ final state has been investigated in \cite{Aad:2021jmg} using full Run 2 data. As before, we here use the model file we used in the whole paper, which corresponds to a UV-complete model, at leading-order. Again the mixing angle has been set to $\sin\theta\,=\,1/\sqrt{2}$. Values are shown in table \ref{tab:bb2}. We use $\slashed{E}_\perp\,\geq\,180\,\GeV,\,|p_{\perp,b}|\,\geq\,50\,\GeV$ for the numbers in that table.
\begin{center}
\begin{table}[tb!]
\begin{center}
\begin{tabular}{c||c||c|c}
$m_a\, [\GeV]$&$\sigma[\pb]$&fac&$\sigma^\text{excl}[\pb]$\\ \hline
60&0.000860 (4)&20&0.01720 (8)\\
100&0.000730 (4)&30&0.0219 (1)\\
200&0.000422 (2)&100&0.0422 (2)\\
300&0.0002336 (8)&140&0.0327 (1)\\
500&0.00003536 (8)&400&0.01414 (3)
\end{tabular}
\caption{\label{tab:bb2} Results for parton-level generation with rudimentary cuts for points which are excluded by $b\bar{b}+\slashed{E}_\perp$ searches in \cite{Aad:2021jmg} via $a$ mediation. Other values are set to $m_H\,=\,m_A\,=\,m_{H^\pm}\,=\,600\,\GeV,\,m_\chi\,=\,1\,\GeV,\,\cos(\be-\al)\,=\,0,\,\lam_3\,=\,\lam_{P_1}\,=\,\lam_{P_2}\,=\,3,\,y_\chi=1,\,\tan\be\,=\,1.$ The factor corresponds to a rough estimate of the multiplicative factor from figure 10 of that reference. The last column gives the actual exclusion cross section.}
\end{center}
\end{table}
\end{center}
This leads to
\begin{\eqn*}
b\bar{b}+\slashed{E}_\perp:\sigma\,\leq\,0.01420\,\pb
\end{\eqn*}

The same channels have been investigated in \cite{Aaboud:2017rzf} using an integrated luminosity of $36\,\fb^{-1}$. In tables \ref{tab:36cms} and \ref{tab:36cmsbb}, we again show the production cross section for selected points from that reference, as well as resulting upper bounds on these cross sections. For $t\bar{t}+\slashed{E}_\perp$, we use $\slashed{E}_\perp\,\gtrsim\,300\,\GeV$; for $b\,\bar{b}+\slashed{E}_\perp$, we take $\slashed{E}_\perp\,\gtrsim\,180\,\GeV,\,p_{\perp,b}\,\gtrsim\,20\,\GeV$. As in the comparison with results from \cite{Aad:2020aob}, we used a slightly different way to generate the event samples for consistency reasons. 
\begin{center}
\begin{table}[tb!]
\begin{center}
\begin{tabular}{c|c||c||c|c}
$m_a$&$m_\chi$&$\sigma[\pb]$&fac&$\sigma^\text{excl}[\pb]$\\ \hline
70\,\GeV&1\,\GeV&0.02536 (8)&1.37&0.0347 (1)\\
100\,\GeV&1\,\GeV&0.02392 (1) &1.41&0.03373 (1)\\
200\,\GeV&1\,\GeV&0.01848 (8)&1.94&0.0359 (2)\\
300\,\GeV&1\,\GeV&0.01292 (8)&2.7&0.0349 (2)\\
500\,\GeV&1\,\GeV&0.00270 (2)&13.65&0.0369 (3)\\
10\,\GeV&30\,\GeV&0.000520 (4) (1)&44.56&0.0232 (2)\\
10\,\GeV&50\,\GeV&0.000368 (4)&63.25&0.0233 (3)\\
10\,\GeV&100\,\GeV&0.000170 (3)&166.52&0.023 (3)\\
10\,\GeV&200\,\GeV&0.000031(4)&736.94&0.0245 (3)\\
10\,\GeV&300\,\GeV&2.92 (4) $\times\,10^{-7}$&2346.43&0.000685 (9)\\
10\,\GeV&500\,\GeV&4.94 (4) $\times\,10^{-8}$&14628.82&0.000723 (5)
\end{tabular}
\caption{\label{tab:36cms} Results for parton-level generation with rudimentary cuts for points which are excluded by $t\bar{t}+\slashed{E}_\perp$ searches in \cite{Aaboud:2017rzf}. Other values are set to $m_H\,=\,m_A\,=\,m_{H^\pm}\,=\,600\,\GeV,\,\cos(\be-\al)\,=\,0,\,\lam_3\,=\,\lam_{P_1}\,=\,\lam_{P_2}\,=\,3,\,y_\chi=1,\,\tan\be\,=\,1.$ The factor has been taken from HEPData \cite{hep1710}. The last column gives the actual exclusion cross section.}
\end{center}
\end{table}
\end{center}
\begin{center}
\begin{table}[tb!]
\begin{center}
\begin{tabular}{c||c||c|c}
$m_a$&$\sigma[\pb]$&fac&$\sigma^\text{excl}[\pb]$\\ \hline
70&0.001808 (8)&316.33&0.572 (3)\\
100&0.001584 (8)&400.243&0.634 (3)\\
200&0.001044 (4)&876.85&0.915 (4)\\
300&0.000479 (2)&1875.75&0.898 (4)\\
500&$6.83 (4) \times\,10^{-5}$&8581.11&0.586 (3)
\end{tabular}
\caption{\label{tab:36cmsbb} Results for parton-level generation with rudimentary cuts for points which are excluded by $b\bar{b}+\slashed{E}_\perp$ searches in \cite{Aaboud:2017rzf}. Other values are set to $m_H\,=\,m_{H^\pm}\,=\,m_A\,=\,600\,\GeV,m_\chi\,=\,1\,\GeV,\,\,\cos(\be-\al)\,=\,0,\,\lam_3\,=\,\lam_{P_1}\,=\,\lam_{P_2}\,=\,3,\,y_\chi=1,\,\tan\be\,=\,1.$ The factor has been taken from HEPData \cite{hep1710}. The last column gives the actual exclusion cross section.}
\end{center}
\end{table}
\end{center}
Following the same logic as before, we now require that
\begin{\eqn*}
\sigma_{t\bar{t}+\slashed{E}_\perp}\,\leq\,0.000703\,\pb;\,\sigma_{b\bar{b}+\slashed{E}_\perp}\,\leq\,0.578\,\pb
\end{\eqn*}
with the cuts as specified above\footnote{Note different cuts apply. Also, in \cite{Aad:2021jmg}, results for $m_a\,<\,m_\chi$ were not presented. These lead to quite strong constraints in the analysis in \cite{Aaboud:2017rzf}.}.
\item{\bf Bounds from monojet searches}\\
We also compare to the results for monojet searches as presented in \cite{CMS:2021far}. As discussed above, this work uses a different model file, which we accomodated for by setting $\sin\theta\,=\,\frac{1}{\sqrt{2}}$ and rescaling our result accordingly. Bounds on the signal strength are taken from Hepdata \cite{hepd20004}. Our results are presented in table \ref{tab:monojet}. We applied a cut on the missing transverse energy $\slashed{E}_\perp\,\gtrsim\,250\,\GeV$.
\begin{center}
\begin{table}[tb!]
\begin{center}
\begin{tabular}{c||c||c|c}
$m_a$&$\sigma[\pb]$&fac&$\sigma^\text{excl}[\pb]$\\ \hline
80&0.346(2)&0.78&0.270(2)\\
100&0.3352(4)&0.76&0.2547(3)\\
200&0.2812(4)&0.69&0.1940(3)\\
300&0.277(2)&0.51&0.141(1)\\
350&0.342(1)&0.36&0.123(4)\\
400&0.184(2)&0.56&0.1028(9)\\
450&0.1144(4)&0.87&0.0995(3)\\
500&0.0728(4)&1.2&0.0874(5)\\
600&0.0362(2)&2.4&0.0868(4)\\
700&0.01832(8)&4.1&0.0751(3)\\
800&0.00962(3)&6.6&0.0636(2)
\end{tabular}
\caption{\label{tab:monojet} Results for parton-level generation with rudimentary cuts for points which are excluded by monojet searches in \cite{CMS:2021far}. Other values are set to $m_H\,=\,m_{H^\pm}\,=\,m_A\,=\,600\,\GeV,m_\chi\,=\,1\,\GeV,\,\,\cos(\be-\al)\,=\,0,\,\lam_3\,=\,\lam_{P_1}\,=\,\lam_{P_2}\,=\,3,\,y_\chi=1,\,\tan\be\,=\,1.$ For $m_a\,\geq\,500\,\GeV$, the heavy scalar masses have been set to $1\,\TeV$. The factor has been taken from HEPData \cite{hepd20004}. The last column gives the actual exclusion cross section.}
\end{center}
\end{table}
\end{center}
The table translates to a maximal allowed cross section of around $\sigma_\text{lim}\,=\,0.0638\,\pb$.

\end{itemize}
\subsection{Dark matter constraints}\label{sec:dmconst}
The THDMa renders a dark matter candidate and is therefore subject to constraints from astrophysical measurements, such as relic density and direct detection. We have made use of MadDM \cite{Backovic:2013dpa,Backovic:2015cra,Ambrogi:2018jqj} for the calculation of relic density. For direct detection, we implemented the analytic expressions presented in \cite{Ipek:2014gua}\footnote{A more detailed analysis, as e.g. presented in \cite{Arcadi:2017wqi,Sanderson:2018lmj,Abe:2018emu,Abe:2019wjw}, is beyond the scope of the current work.}. We compare these values to limits from the Planck collaboration \cite{Planck:2018vyg}, and require that
\begin{\eqn}\label{eq:omup}
\Omega\,h^2\,\leq\,0.1224
\end{\eqn}
which corresponds to a 2 $\sigma$ limit. Direct detection bounds are compared to maximal cross section values $\sigma^\text{Xenon1T}_\text{max}\,\lb m_{\chi}\rb$  using XENON1T result\cite{Aprile:2018dbl}, which we implemented in terms of an approximation function\footnote{The numerical values have been obtained using the Phenodata database \cite{PhenoData}.}. Note we rescale bounds from direct detection according to
\begin{\eqn}\label{eq:resc}
\sigma_\text{max}\,\lb m_{\chi,i},\Omega_i \rb\,=\,\sigma^\text{Xenon1T}_\text{max}\,\lb m_{\chi}\rb\,\frac{0.1224}{\Omega_i},
\end{\eqn}
where $m_{\chi,i},\,\Omega_i$ refer to the dark matter and relic density of the specific parameter point $i$ tested here.
\section{Scan setup and results}\label{sec:scan}
\subsection{Scan setup}
The above constraints are implemented using an interplay of private and publicly available codes in several steps. We discuss these in the following. This also means that, as bounds are implemented successively, subsequent constraints are imposed on the subset of parameter points which have passed previous constraints.
\begin{enumerate}
\item{}In a first step, we impose perturbativity, positivity, and perturbative unitarity constraints as discussed in section \ref{sec:thconst}. We also set a lower limit on $\tan\be$ from the fit function for $B\,\rightarrow\,X_s\gamma$ derived in \cite{Misiak:2020vlo,mm}.
\item{}Points passing these bounds are then passed on to Spheno, which is used for the calculation of limits from electroweak oblique parameters, $B_s\,\rightarrow\,\mu^+\mu^-,\,\Delta\,M_s$, as well as the total widths for all particles. These values are confronted with the experimental constraints discussed in section \ref{sec:ewetcconst}.
\item{}The remaining parameter points are now passed through HiggsBounds and HiggsSignals, for comparison with null-results from experimental searches as well as Higgs signal strength measurements which are contained in these tools.
\item{}We then calculate the relic density as well as direct detection constraints, which are discussed in section \ref{sec:dmconst}. 
\item{}Parameter points which have passed all remaining constraints are confronted with the collider bounds presented in section \ref{sec:collconst}.
\end{enumerate}

Our initial scan ranges are determined by a number of prescans to determine regions of parameter space that are highly populated, while still rendering relavitely low masses up to 1 \TeV. In particular, we set 
\begin{eqnarray}\label{eq:ranges}
&&\sin\theta\,\in\,\left[-1;0.8\right];\;\cos\lb \be-\al\rb\,\in\left[ -0.08;0.1 \right];\,\tan\be\,\in\,\left[0.52;9\right], \nonumber\\
&&m_H\,\in\,\left[500;1000\right]\GeV,\,m_A\,\in\,\left[600; 1000 \right]\GeV,\,m_{H^\pm}\,\in\left[800;1000\right]\GeV, \nonumber\\
&&m_a\,\in\,\left[0;m_A\right],\,m_\chi\,\in\,\left[0,m_a\right], \nonumber\\
&&y_\chi\,\in\,\left[-\pi;\pi\right],\,\lambda_{P_1}\,\in\,[0;10],\,\lambda_{P_2}\,\in\,\left[0;4\,\pi\right],\,\lambda_3\,\in\left[-2;4\,\pi\right].
\end{eqnarray}
The values of $m_h\,=\,125\,\GeV$ and $v\,=\,246\,\GeV$ are set according to measurements of the Higgs boson mass as well as precision measurements. Note the above choices do not imply that values outside the above regions are stricly forbidden (apart from bounds we will discuss explicitely below), but these were found to render a relatively large acceptance rate for the scan discussed above. We furthermore apply the following relations between masses 
\begin{eqnarray*}
m_A-m_H\,\in\,\left[-300;400\right]\GeV,&m_{H^\pm}-m_H\,\in\,\left[-150;300\right]\GeV,& m_A-m_{H^\pm}\,\in\left[-200;200  \right]\,\GeV,\\m_a\,< \,m_H+400\,\GeV,&
m_a< m_{H^\pm}+200\,\GeV,& m_\chi<m_H-200\,\GeV,\\m_\chi\,<\,m_A-300\,\GeV, &m_\chi<m_{H^\pm}-400\,\GeV\,&m_\chi<m_a/2+50\,\GeV,
\end{eqnarray*}
which were superimposed on the original scan ranges\footnote{We obviously adjusted ranges such that all $m_i\,\geq\,0\,\GeV$.}. Note that a priori also points outside the above regions were allowed; these were chosen to optimize the selections process imposed by cuts.

Note that the fit for $B\,\rightarrow\,X_s\,\gamma$ implies a lower bound on $m_{H^\pm}$ of $\sim\,800\,\GeV$\footnote{We are aware that using these observables in a fit rather than as hard bounds might weaken some of the constraints discussed here, as e.g. discussed in \cite{Eberhardt:2020dat} for a normal THDM. This is however beyond the scope of the current work.}.
\subsection{Scan results}
In the following, we discuss the resulting constraints on the parameter space of the THDMa. Note that not all bounds discussed above lead to a direct limit in a two-dimensional parameter plane. In case the effects can be displayed in such a simple way, we discuss them explicitely.

\begin{center}
\begin{figure}[tb]
\begin{center}
\includegraphics[width=0.7\textwidth]{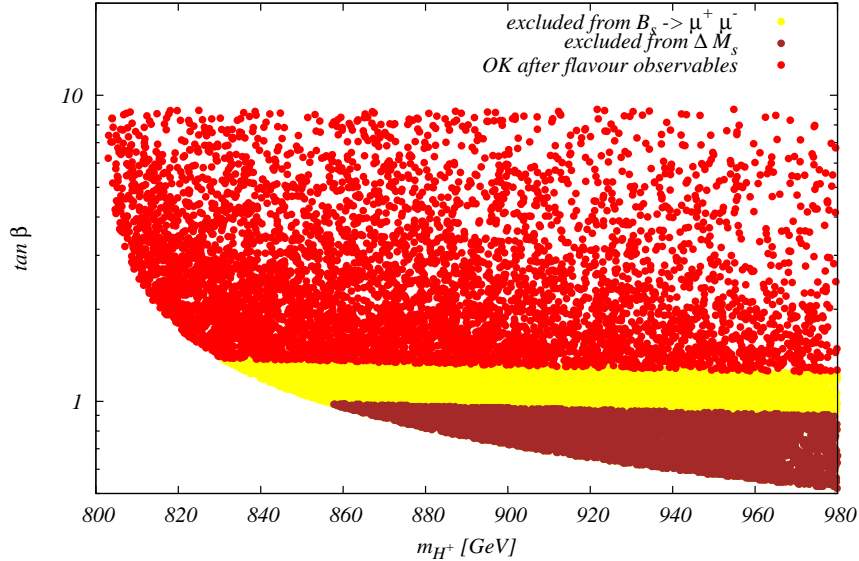}
\caption{\label{fig:brsg} Exclusion in the $\lb m_{H^\pm},\tan\be\rb$ plane after applying flavour constraints. The lower bound for $m_{H^\pm}\,\leq\,850\,\GeV$ is set by the bound on $B\,\rightarrow\,X_s\,\gamma$ as discussed in section \ref{sec:ewetcconst}. }
\end{center}
\end{figure}
\end{center}

\begin{center}
\begin{figure}
\begin{center}
\includegraphics[width=0.45\textwidth]{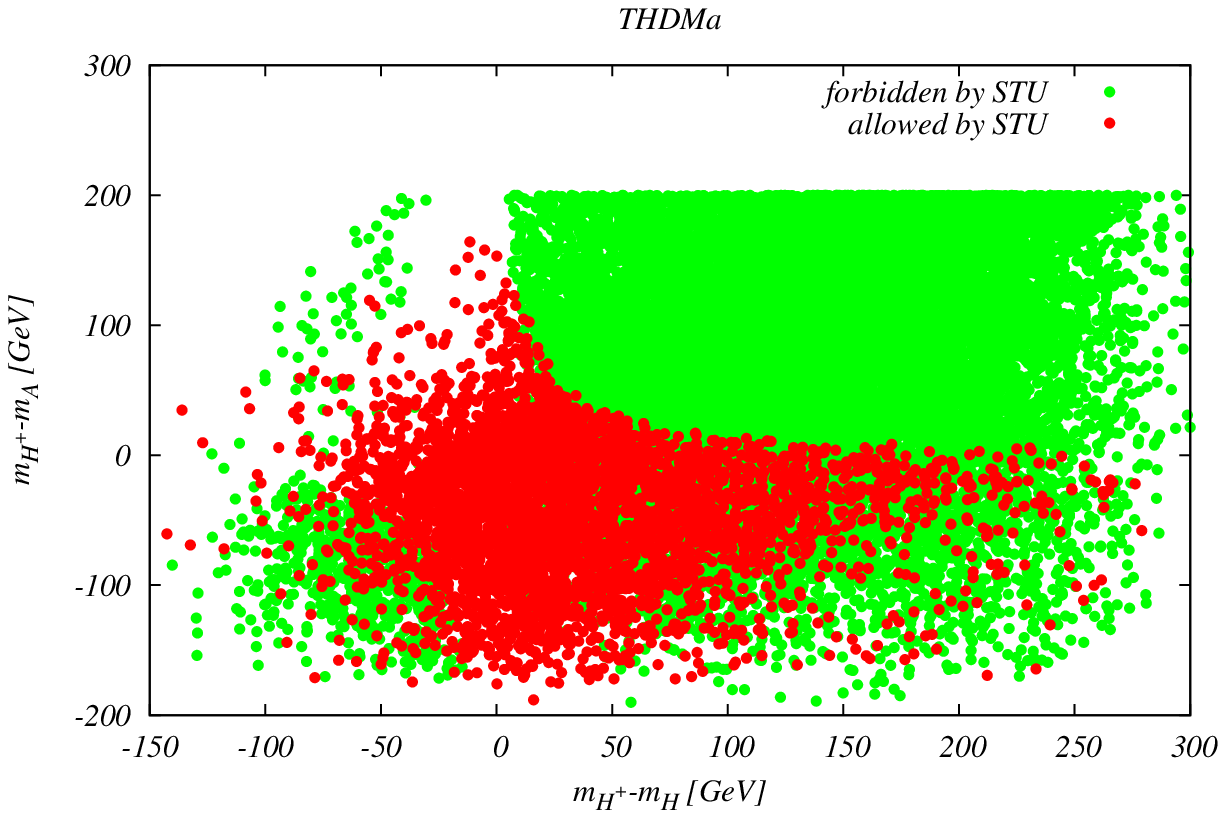}
\includegraphics[width=0.45\textwidth]{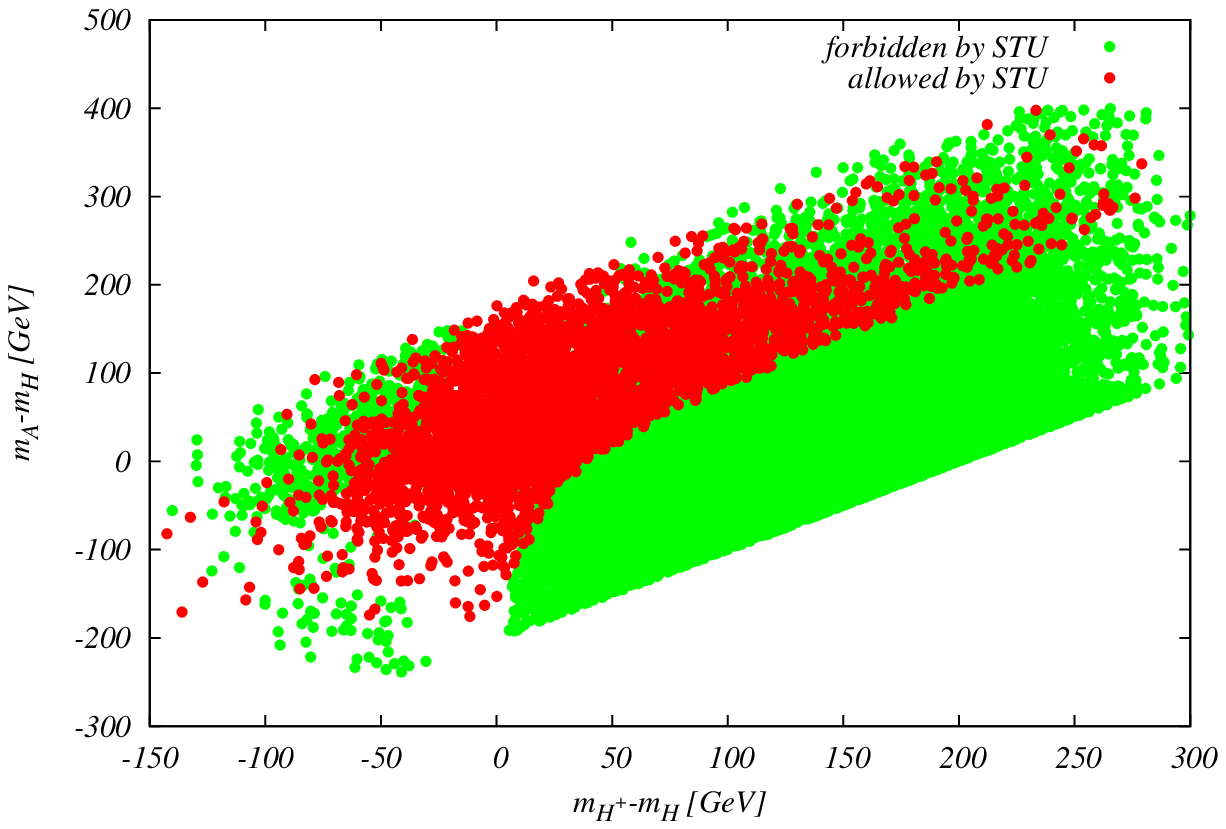}
\caption{\label{fig:stuexcl} Exclusions in the $\lb m_{H^\pm}-m_H,m_{H^\pm}-m_A\rb$ {\sl (left)} and $\lb m_{H^\pm}-m_H,m_{H}-m_A\rb$ {\sl (right)} planes from oblique parameters. We see that regions where both displayed mass differences are large are excluded by the oblique parameters.  }
\end{center}
\end{figure}
\end{center}

\begin{center}
\begin{figure}
\begin{center}
\includegraphics[width=0.7\textwidth]{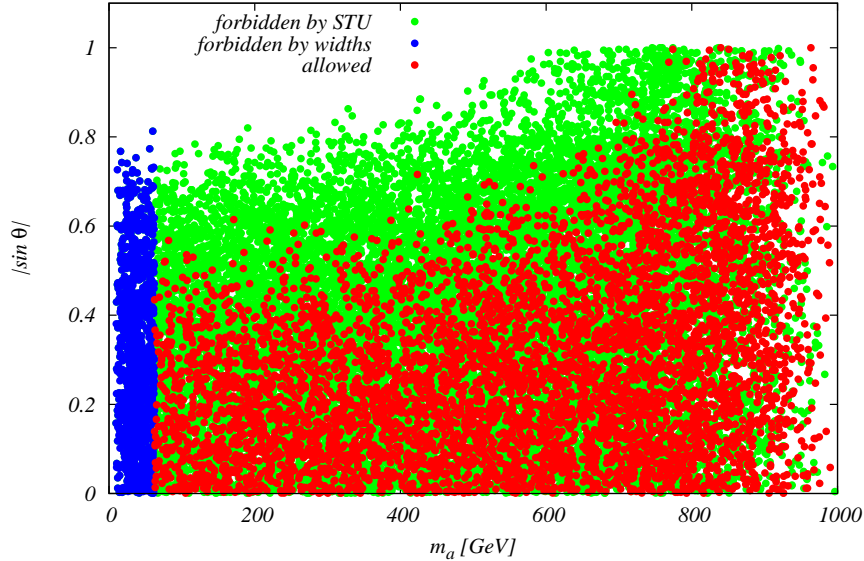}
\caption{\label{fig:stu3} Exclusion in the $\lb m_a,|\sin\theta| \rb$ plane from oblique parameters. Limits from setting an upper bound on the total width of the 125 \GeV~ scalar (eqn. (\ref{eq:125w})) are also shown. }
\end{center}
\end{figure}
\end{center}

\begin{center}
\begin{figure}
\begin{center}
\includegraphics[width=0.45\textwidth]{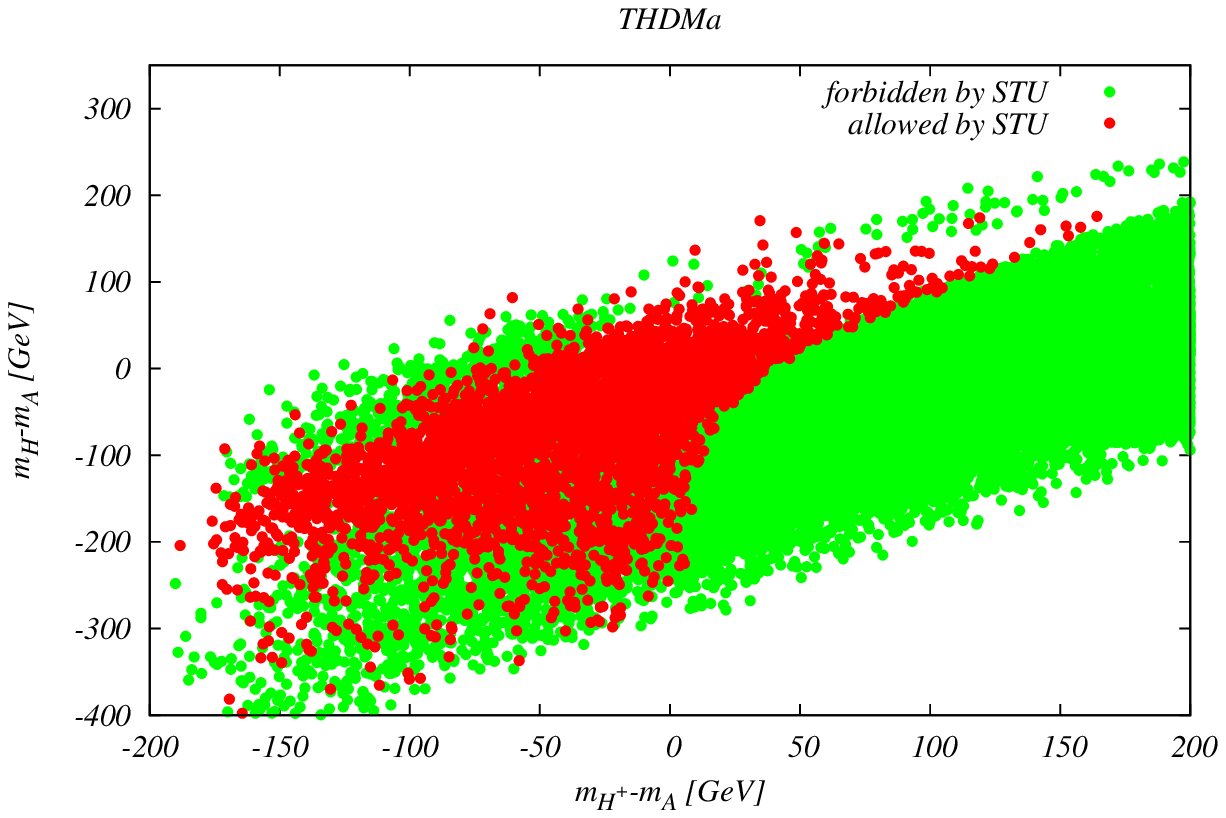}
\includegraphics[width=0.45\textwidth]{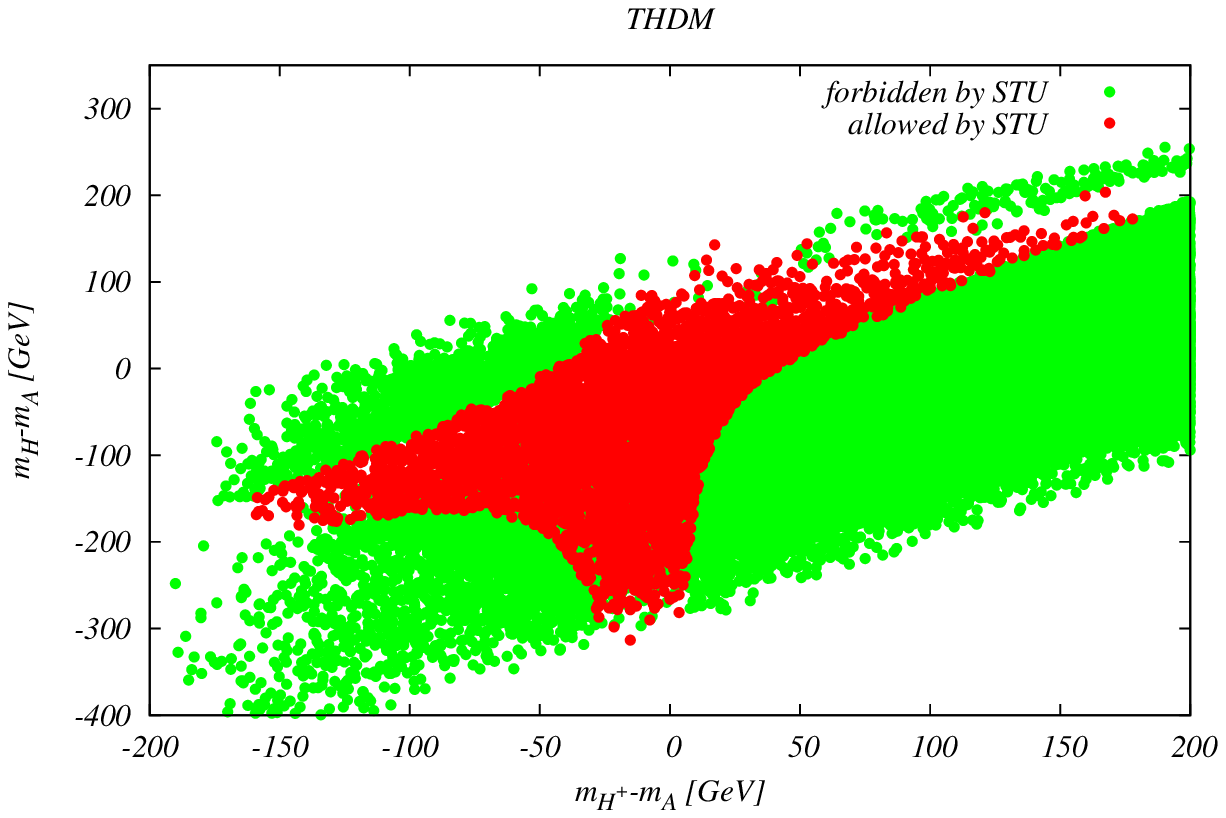}
\caption{\label{fig:stucomp} Exclusions in the $\lb m_{H^\pm}-m_A,m_{H}-m_A\rb$ plane in the THDMa {\sl (left)} and THDM {\sl (right)} from oblique parameters; for the latter, $\sin\theta\,=\,\lam_{P_1}\,=\,\lam_{P_2}\,=\,0$. The admixture of $a$ releases the bounds, as expected.}
\end{center}
\end{figure}
\end{center}

We generate $\sim\,30000$ parameter points in the first step of the above scan. Application of bounds in step 2 includes e.g. constraints from flavour-physics in the $m_{H^\pm},\tan\be$ plane. Results from flavour constraints in the $m_{H^\pm},\tan\be$ plane are shown in figure \ref{fig:brsg}, where similar results have been obtained e.g. in \cite{Enomoto:2015wbn,Arbey:2017gmh,Haller:2018nnx}\footnote{Note the newest result from the LHC experimental combination for $B_s\,\rightarrow\,\mu^+\mu^-$ \cite{ATLAS-CONF-2020-049} increases the discrepancy between experimental result and theory prediction, leading to a larger exclusion region. In the mass range considered here, we found the limit can be approximated by requiring $\tan\be\,\geq\,2.15302-0.000930233\,\frac{m_{H^\pm}}{\GeV}$. Furthermore, in \cite{Haller:2018nnx}, the $\Delta M_s$ bound is not shown in the figure for flavour constraints. However, similar results are obtained as well. TR wants to thank the GFitter authors for clarifying this point.}
After the bounds imposed by theory, the only constraints we find from widths are points which violate (\ref{eq:125w}). This basically rules out all points with $m_a\,\leq\,62.5\,\GeV$, as the $h\,\rightarrow\,a\,a$ decay becomes dominant and can lead to large partial widths \footnote{Note that the respective coupling does not approach zero for $\sin\theta\,\rightarrow\,0$, but is mediated via $\lam_{P_1}$ and $\lam_{P_2}$, cf. e.g. the explicit form of the coupling given in \cite{Bauer:2017ota}.}. All other mass ratios are $\Gamma/m\,\lesssim\,28\%$ for all points that fulfill theory constraints, and maximally around $24\%$ for $\Gamma_A/m_A$ after width bounds are included. As an example, the point with the largest allowed ratio features a dominant branching ratio for $A\,\rightarrow\,h a$, mediated via relatively large $\tan\be\,\sim\,8.1,\,\lam_{P_2}\,\sim\,11.6$ values and masses $m_A\,\,=\,847\,\GeV,\,m_a\,=\,298\,\GeV$. For large values of $\Gamma_H/m_H$, typically $H\,\rightarrow\,a\,Z$ and $H\,\rightarrow\,t\,\bar{t}$ are dominant, with $m_H\,\gtrsim\,800\,\GeV,\,|\sin\theta|\,\sim 0.35-0.65$. Large widths for $a$ typically stem from large $|y_\chi|\,\sim\,1.8$ and small $|\sin\theta|\,\lesssim\,0.2$ values, corresponding to the coupling determining the $a\,\rightarrow\,\chi\,\bar{\chi}$ decay \cite{Bauer:2017ota}. Electroweak precision observables, on the other hand, have an impact on the allowed mass splittings, such that for example regions with $|m_{H^\pm}-m_A|\,\gtrsim\,190\,\GeV$ are basically forbidden. In fact, we see largest deviations in the oblique parameters trace back to too large absolute values of the T-parameter, i.e $T\,\lesssim\,-0.03$, where largest values stem from $|m_{H^\pm}-m_H|\,\gtrsim\,250\,\GeV$. Such deviations typically stem from new physics contributions to gauge-boson propagators, therefore being sensitive to mass differences. As an example, we show the exclusion in the $\lb m_{H^\pm}-m_H, m_{H^\pm}-m_A\rb$ and $\lb m_{H^\pm}-m_H, m_{H}-m_A\rb$  planes in figure \ref{fig:stuexcl}. Another interesting parameter plane is given by $\lb m_a,\,|\sin\theta|\rb$, where the oblique parameters also set a clear limit, see figure \ref{fig:stu3}.  We also investigate how our results compare to the ones obtained in the THDM limit, where $a$ is decoupled. This can be achieved setting $\sin\theta\,=\,\lam_{P_1}\,=\,\lam_{P_2}\,=\,0$. We then compare the limits in the $\lb m_{H^\pm}-m_A, m_{H}-m_A\rb$ plane, cf. fig. \ref{fig:stucomp}. We see that the admixture of $a$ leads to much weaker constraints on $m_{H^\pm}-m_A$.  For the remaining constraints discussed in section \ref{sec:ewetcconst}, the identification of additional regions in two-dimensional parameter planes that are primarily affected is not straightforward.
We now turn to the effects of applying collider constraints via HiggsBounds/ HiggsSignals. As expected in THDMs and its extensions, only narrow stripes around $\cos\lb \be-\al\rb\,=\,0$ remain in agreement with current signal strength measurements (see e.g. \cite{Sirunyan:2018koj,Aad:2019mbh}). 
\begin{center}
\begin{figure}
\begin{center}
\includegraphics[width=0.7\textwidth]{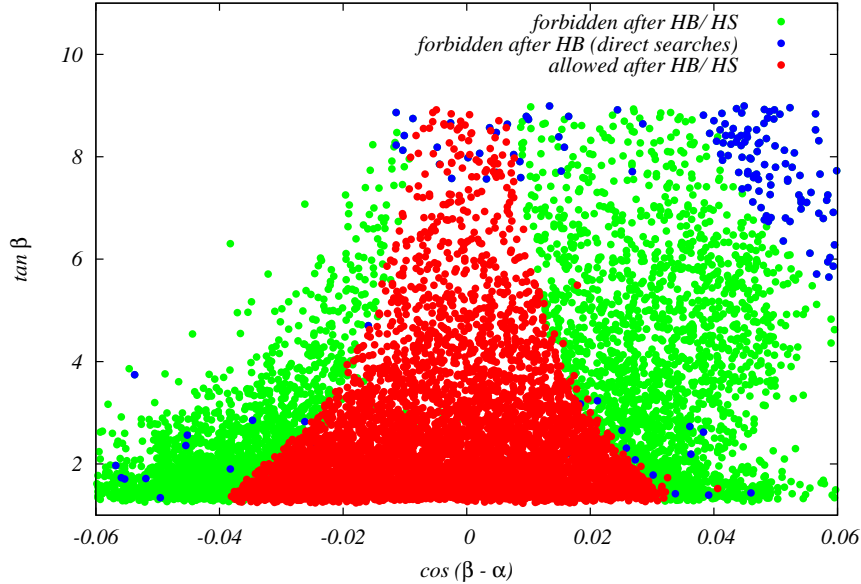}
\caption{\label{fig:hbhs} Exclusion in the $\cos(\be-\al),\tan\be$ plane after HiggsBounds (HB) and HiggsSignals (HS). }
\end{center}
\end{figure}
\end{center}
Using the tools described above, we find the allowed/ forbidden areas in the $\lb \cos\lb \be-\al\rb,\tan\be\rb$ plane as shown in figure \ref{fig:hbhs}\footnote{Comparing to the results presented e.g. in \cite{Kling:2020hmi,Han:2020lta}, we see that constraints for negative values of $\cos(\be-\al)$ agree, while we find more stringent constraints for positive values. However, note that we use different approaches, and for example HiggsSignals makes use of the most recent constraints as well as STXS information \cite{ATLAS-CONF-2020-027}. We thank the authors of \cite{Kling:2020hmi} for useful correspondence concerning this point.}. Some points are additionally ruled out by  $H/ a\,\rightarrow\,\tau\,\tau$ \cite{Aad:2020zxo}, $H\,\rightarrow\,h_{125}h_{125}$ \cite{Aaboud:2018knk} and $H\,\rightarrow\,a\, Z$ \cite{Aaboud:2018eoy,ATLAS-CONF-2020-043} searches. Values of $\cos\lb \be-\al\rb\,>\,0.04$ and $\tan\be\,\gtrsim\,5$ are excluded from $h_{125}\,\rightarrow\,Z\,Z$ \cite{Chatrchyan:2013mxa}. Points which are still allowed after the application of HiggsBounds and HiggsSignals are outside the region sensitive to the full Run 2 $A\,\rightarrow\,H\,Z$ search \cite{Aad:2020ncx}, assuming on-shell production and decay\footnote{We tested this by comparing to cross sections for $p\,p\,\rightarrow\,A\,\rightarrow\,Z\,H$ production and successive decays $H\,\rightarrow\,b\,\bar{b}$ from figure 10 in that reference.}.  Furthermore, the region where $\cos\lb \be-\al\rb\,\lesssim\,-0.05,\,\tan\be\,\gtrsim\,5$ is mainly ruled out from both perturbative unitarity and perturbativity.

\begin{center}
\begin{figure}
\begin{center}
\includegraphics[width=0.45\textwidth]{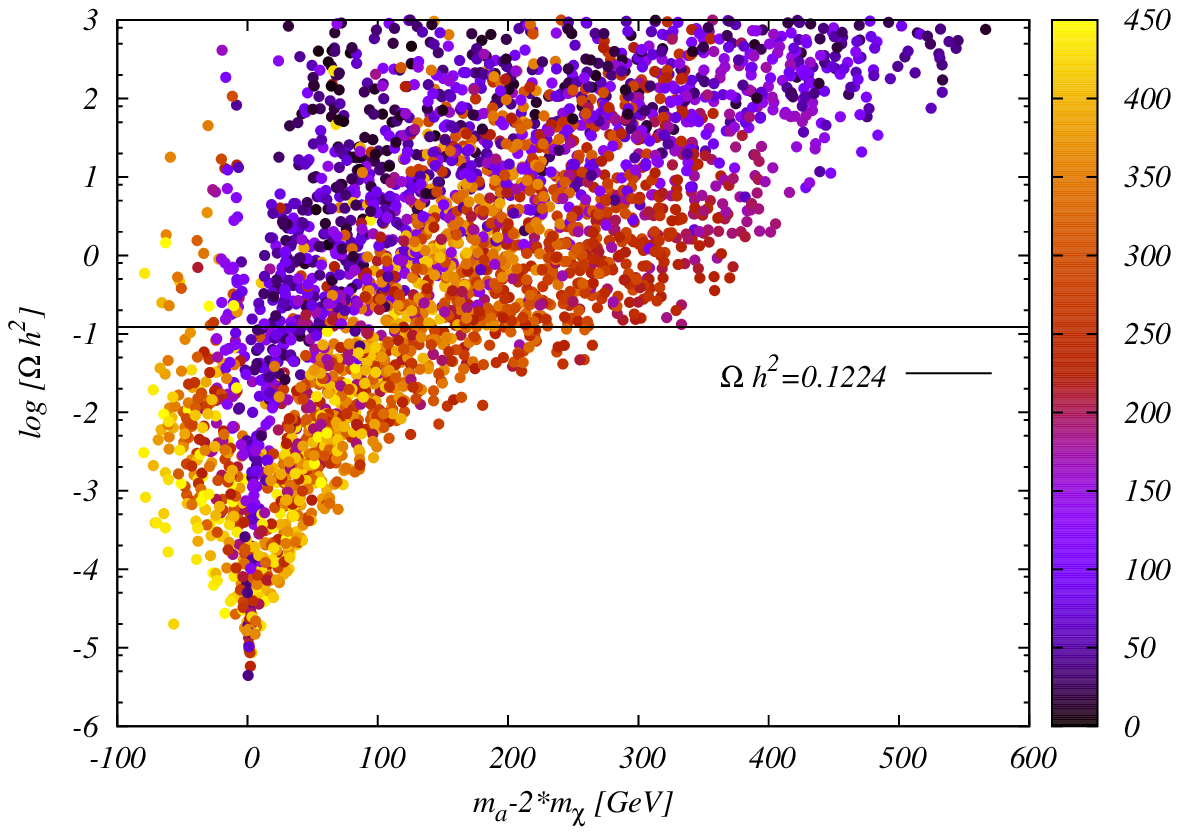}
\includegraphics[width=0.45\textwidth]{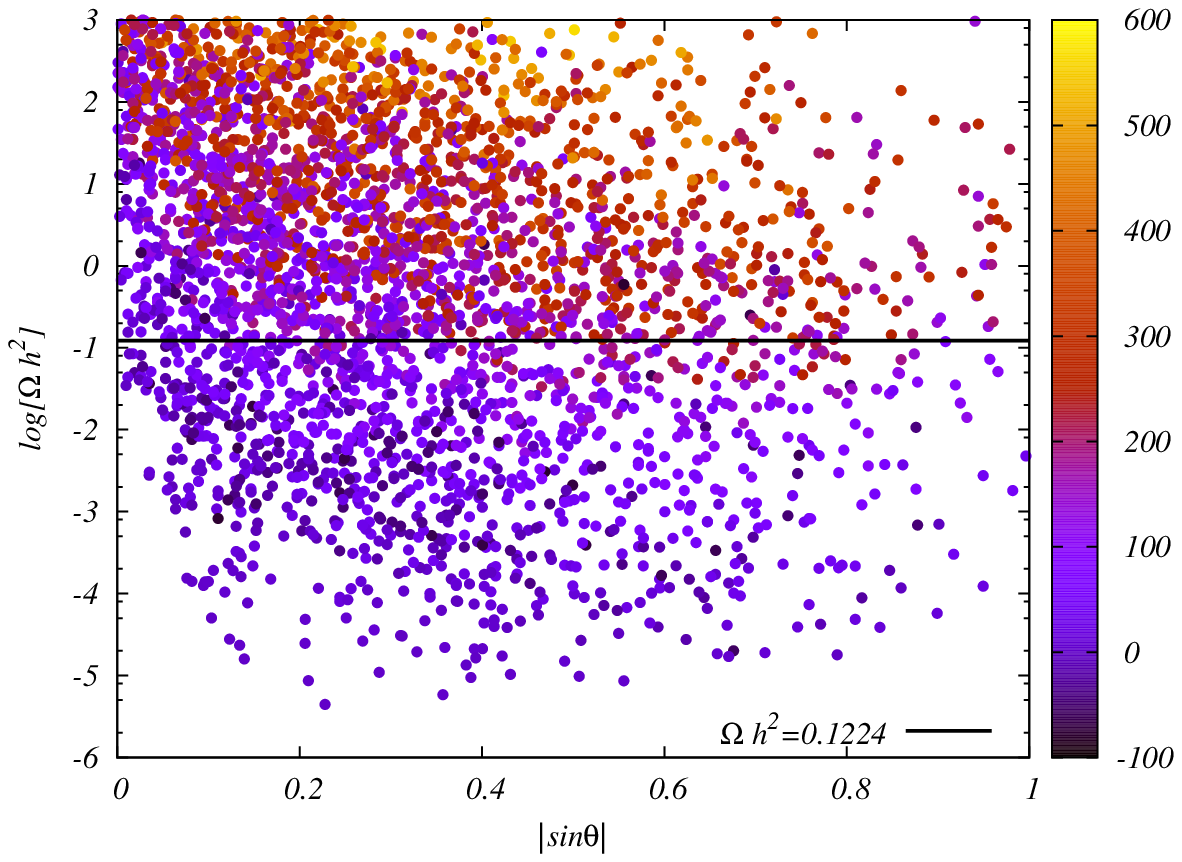}
\caption{\label{fig:om} {\sl Left:} 
Relic density as a function of $m_a-2\,m_\chi$, with the color coding referring to the mass of the DM candidate. Heavier $m_\chi$ tend to lead to smaller values of the relic density.  {\sl Right:} Relic density as a function of $|\sin\theta|$, with the color coding referring to the mass difference $m_a-2\,m_\chi$. Lower values of $|\sin\theta|$ lead to smaller annihilation cross sections and therefore larger values of the relic density.}
\end{center}
\end{figure}
\end{center}

Constraints from relic density, i.e. the requirement that eqn. (\ref{eq:omup}) is fulfilled, depends on many parameters. In general, however, small values of $|m_a\,-\,2\,m_\chi|$ can trigger efficient annihilation processes via the resonant s-channel diagram, and lead to relic density values between $10^{-6}$ and the upper bound. For larger mass differences $\mathcal{O}\lb 200\, \GeV\rb$ or larger, basically all points are excluded. Dominant annihilation channels are $\chi \bar{\chi}\,\rightarrow\,b\,\bar{b}$ and $\chi \bar{\chi}\,\rightarrow\,t\,\bar{t}$, where the latter channel opens up for $m_a\,\gtrsim\,2\,m_{t}$ and tends to lead to smaller values of relic density. We show the values of relic density as a function of the above mass difference in figure \ref{fig:om}.
Similarly, for $\sin\theta\,\rightarrow\,0$ annihilation cross sections tend to become small, such that relic density predictions exceed the value measured by the Planck collaboration, cf. eq. (\ref{eq:omup}). We display the dependence on this variable in figure \ref{fig:om}, where in addition color coding indicates the mass difference $m_a\,-\,2\,m_\chi$.
Considering only points that pass the relic density constraints, we found that 
direct detection does not impose any additional constraints. {A more detailed analysis, as e.g. presented in \cite{Arcadi:2017wqi,Sanderson:2018lmj,Abe:2018emu,Abe:2019wjw}, could modify this statement. However, from figure 24 in \cite{Abe:2018bpo}, we see that these constraints are effective only for masses $m_a\,\lesssim\,60\,\GeV$ and additionally depend on the mixing angle. We have verified that the few points that feature this mass range are not excluded when comparing to the above figure. Further comments on the sensitivity of future direct detection experiments for this model can e.g. be found in \cite{Abe:2019wjw}. 

Before discussing constraints from direct searches, we will briefly summarize the effects the bounds considered above have imposed on the original parameter space
\begin{itemize}
\item{}B-physics constraints set a lower bound on $\tan\be$ as a function of $m_{H^\pm}$; in general, $\tan\be\,>\,1$.
\item{}Signal strength measurements reduced the available parameter space for $\cos\lb\be-\al\rb$, such that now $\cos\lb\be-\al\rb\,\in\left[ -0.04;0.04\right]$.
\item{}Relic density reduced the available parameter space to regions where $m_a\,-\,2\,m_\chi\,\in\left[-100;300 \right]\,\GeV$.
\item{}Furthermore, oblique parameters reduce the allowed mass differences, especially in the THDM scalar sector.
\item{}In the general scan, also the upper limits of $m_a$ in dependence on $m_H,\,m_{H^\pm}$ are reduced by more than 100 \GeV. However, this is an artefact of the scan, and one can easily repopulate these regions in a more finetuned setup.
\item{}We also see that $m_{H}\,>\,550\,\GeV,\,m_A\,>\,750\,\GeV$. In both cases, a large number of points are excluded from $S,T,U$ observables, as these variables favour small mass differences between the new scalars. If we force $m_A\,\leq\,750\,\GeV$, for example, out of 1000 points only one survives which has very degenerate $m_H,\,m_{H^\pm}$ masses, differing on the permill level. If we force $m_H\,\leq\,550\,\GeV$, more points survive, which feature a small $m_{H^\pm},m_A$ difference, typically $\,\lesssim\,10\%$ difference, as well as small mass differences $|m_a\,-\,2\,m_\chi|\,\lesssim\,120\,\GeV$. 
In a general, non-finetuned scan, the lower bound on $m_{H^\pm}$ forces the mass scale of $H,\,A$ to be $\gtrsim\,530-550\,\GeV$.
\end{itemize}
All other parameters still populate the original regions set in eqn. (\ref{eq:ranges}).

Regarding direct collider constraints as discussed in section \ref{sec:collconst}, we find the following results

\begin{itemize}

\item{}The strongest constraints stem from $h+\slashed{E}_\perp$ searches presented in \cite{ATLAS:2021shl}, cutting out around $\sim\,28\%$ of the leftover parameter space. As we let all parameters float freely, limits in 2-dimensional planes are not straightforward. Using our approach, only points for which $m_a-2\,m_\chi\,\geq\,0$ are affected. Furthermore, points where $m_\chi\,\geq\,400\,\GeV$ are not affected. This however is e.g. due to a strong correlation between $m_\chi$ and $m_a$, which in turn influences the mass difference $m_A-m_a$ and the $A\,\rightarrow\,h\,a$ branching ratio.
\item{}The $\ell\ell+\slashed{E}_\perp$ searches cut out roughly $9\%$ of the available parameter space. As in \cite{Sirunyan:2020fwm}, regions with $m_a\,\lesssim\,500\,\GeV$ are mainly affected. However, as we scan over all parameters, also many points for which $m_a\,\lesssim\,500\,\GeV$ are still allowed. In addition, the branching ratio for $H\,\rightarrow\,a\,Z$ swiftly goes to 0 when $m_H-m_a\,\rightarrow\,m_Z$, and only has reasonably large values when $m_H-m_a\,\gtrsim\,200\,\GeV$, so that only points obeying this mass hierarchy are excluded. Similarly, as the above branching ratio $\sim\,\sin^2\,\theta$, we found that points for which $|\sin\theta|\,\lesssim\,0.15$ are not excluded.
\item{}The bounds from the $p\,p\,\rightarrow\,H^+\,\bar{t}\,b,\,H^+\,\rightarrow\,\bar{b}{t}$ searches \cite{Aad:2020kep} do not constrain the parameter space here, as production cross sections for the process $p\,p\,\rightarrow\,H^+\,\bar{t}\,b$ are $\mathcal{O}\lb 10^{-2}\, \pb\rb$. This can be traced back to the lower limit on $\tan\be$ from B-physics observables, cf. fig \ref{fig:brsg}. The same holds for searches in the same production mode, but with different final states \cite{ATLAS:2021upq}.
\item{}Searches for $t\,\bar{t}+\slashed{E}_\perp$ cut out a relatively small region of parameter space,$\sim\,1\,\%$, in the range where $m_a\,\lesssim\,300\,\GeV,\,m_a-2\,m_\chi\,\in\,\left[0;\,40\right]\,\GeV,\,\tan\be\,\lesssim\,2$. As before, several parameters contribute, so one also finds allowed points within the regions mentioned above. All of these are also subject to at least one other constraint from the searches mentioned above.
\item{}None  of the points is excluded by the $W^+\bar{t}+\slashed{E}_\perp/ W^-t+\slashed{E}_\perp,\,b\,\bar{b}+\slashed{E}_\perp$ and monojet searches considered here.
\end{itemize}

\section{Predictions for $e^+e^-$ colliders}\label{sec:e+e-}
Out of the remaining parameter space, we now investigate the magnitude of expected rates at possible future $e^+e^-$ colliders. In the limit where $\sin\theta\,\rightarrow\,0$, we recover the decoupling scenario of a standard THDM. It is therefore interesting whether we can find regions in parameter space where novel signatures, and explicitely final states with missing energy, give the largest rates.

If we focus on the neutral sector and pair-production processes, possible final states are given by $ha,\,hA,\,HA,\,Ha$. For the first two processes, the sum of masses remains $\lesssim\,1\,\TeV$, so colliders with a center-of-mass energy in that range might be relevant. For the latter two, on the other hand, masses can range above $1\,\TeV$, and we can consider cross sections at a 3 \TeV~collider.

Note that $h\,\left[ H\right]\,A$ production in a THDM is mediated mainly via Z-boson exchange in the s-channel, where the corresponding coupling is proportional to $\cos\,\lb \be -\al\rb\,\left[ \sin\lb \be-\al\rb\right]$ respectively. As $\cos\lb\be-\al\rb$ is heavily constrained, the corresponding cross-sections are quite small, reaching $\mathcal{O}\lb 10^{-3}\fb\rb$. Therefore, we will here concentrate on $HA\,(a)$ production and decay. For this channel, we find that at a 3 \TeV~collider, production cross-sections can reach up to $1\,\fb$, where largest cross section values are achieved for $m_A+m_H\,\sim\,1400\,\GeV$. Dominant decay modes for such points are displayed in figure \ref{fig:brsAH}.
\begin{figure}
\begin{center}
\includegraphics[width=0.5\textwidth]{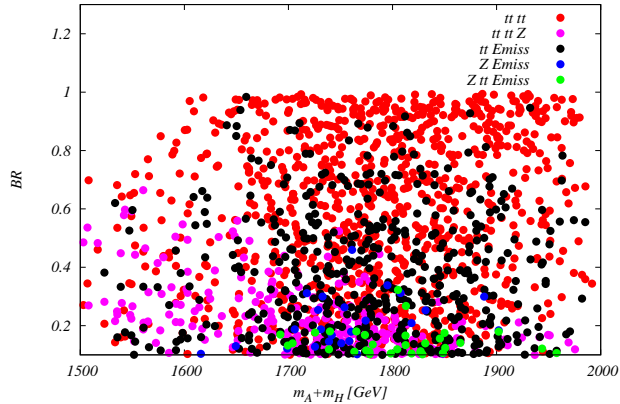}
\caption{\label{fig:brsAH} Combined branching ratios for $HA$ final states, as a function of the mass sum.}
\end{center}
\end{figure}
Decays without missing energy lead to $t\bar{t}t\bar{t}$ and $t\,\bar{t}\,t\,\bar{t}\,Z$ final states, where the first in dominant in large regions of parameter space. The first decay mode that is novel with respect to standard THDMs is the $t\,\bar{t}+\slashed{E}$ final state. In order to identify regions where this state dominates, we show the expected $t\bar{t}t\bar{t}$ and $t\bar{t}+\slashed{E}$ cross sections in figure \ref{fig:xsecs}, where in the right plot we additionally include contributions mediated via $Ha$ production.
\begin{figure}
\includegraphics[width=0.49\textwidth]{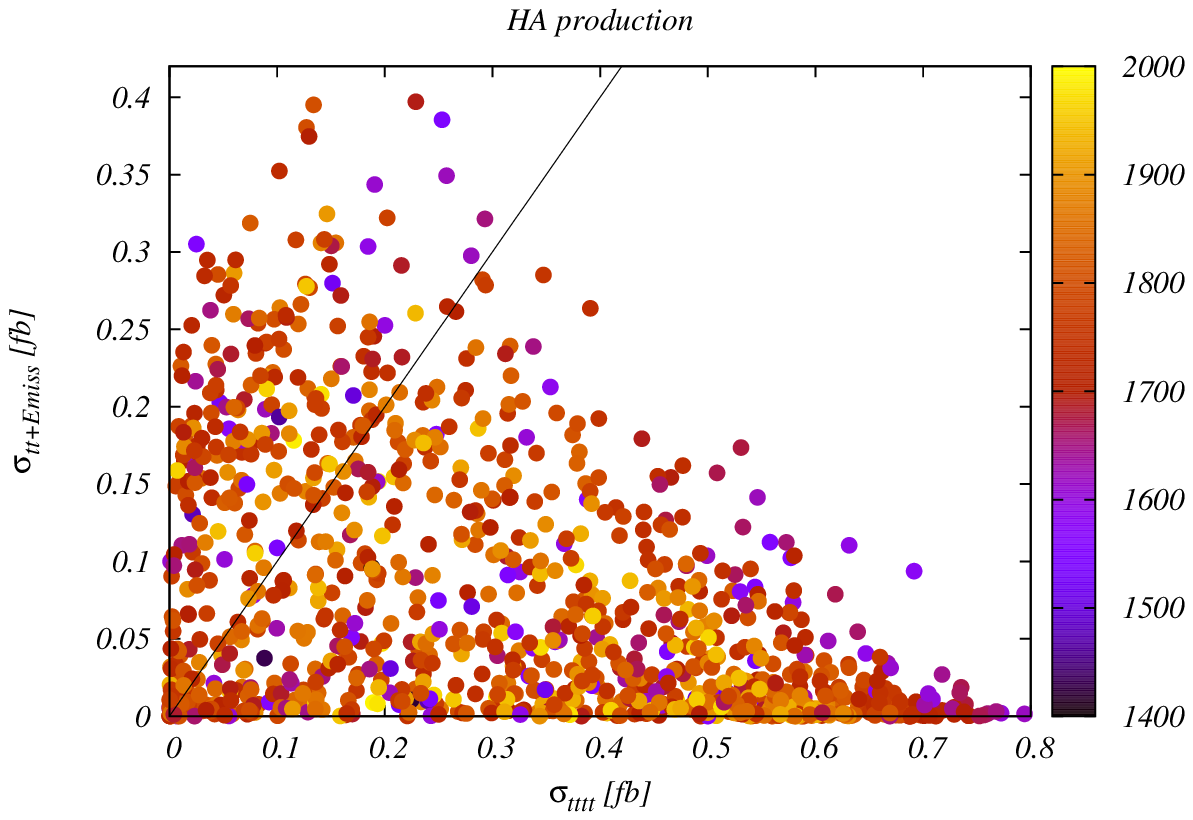}
\includegraphics[width=0.49\textwidth]{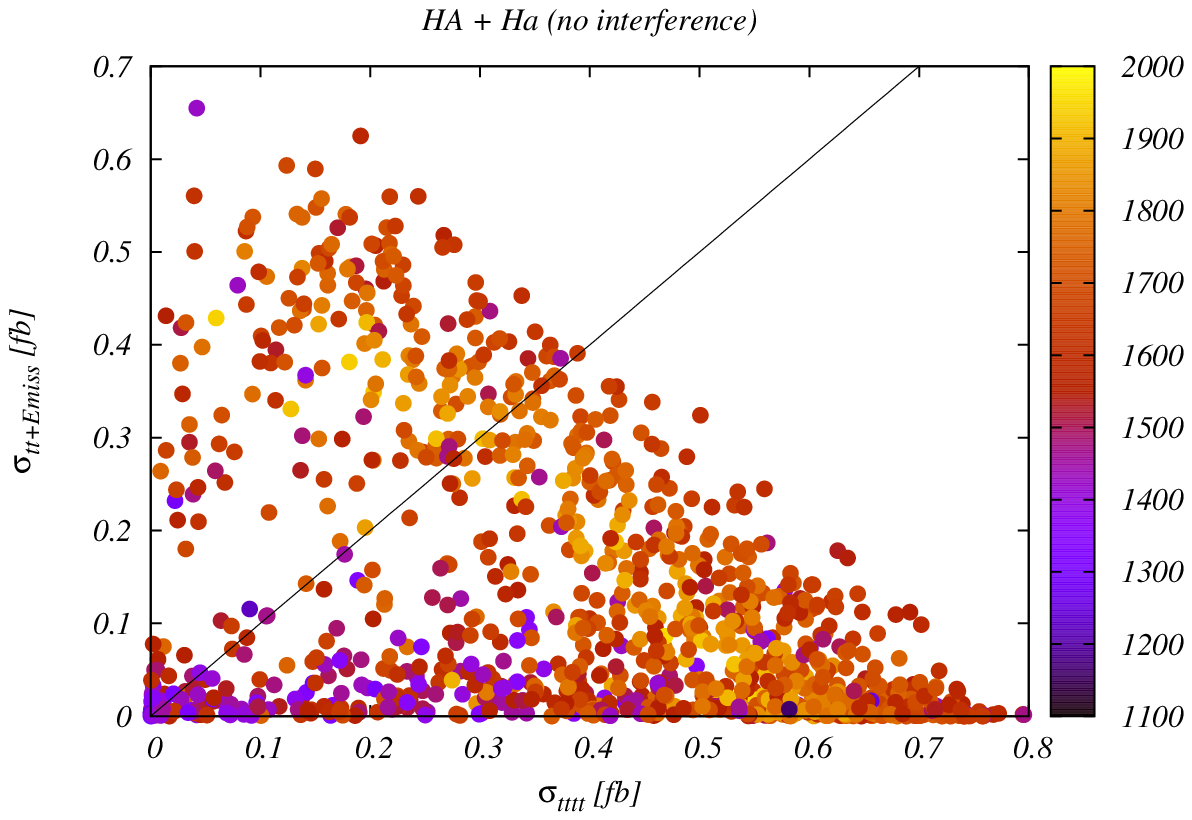}
\caption{\label{fig:xsecs} Production cross sections for $t\bar{t}t\bar{t}$ and $t\bar{t}+\slashed{E}$ final states, using a factorized approach. {\sl (Left:)} via $HA$ and {\sl (right:)} via $HA+Ha$ production. Color coding refers to the mass scale, which is defined as $m_H+m_A$ {\sl (left)}/  $m_H+0.5\,\times\lb m_A+m_a \rb$ {\sl (right)}, respectively.}
\end{figure}
We see that indeed we can identify regions where $t\bar{t}+\slashed{E}$ dominates and renders the largest rates. As an example, we here list the parameters of the "best point":
\begin{eqnarray}
&&\sin\theta\,=\,-0.626,\;\cos\lb \be-\al\rb\,=\,0.0027,\,\tan\be\,=\,3.55 \nonumber\\
&&m_H\,=\,643\,\GeV,\,m_A\,=\,907\,\GeV,\,m_{H^\pm}\,=\,814\,\GeV, \nonumber\\
&&m_a\,\,=\,653\,\GeV,\,m_\chi\,=\,277\,\GeV, \nonumber\\
&&y_\chi\,\,=\,-1.73,\,\lambda_{P_1}\,=\,0.18,\,\lambda_{P_2}\,=\,2.98,\,\lambda_3\,=\,8.63.
\end{eqnarray}
The total widths are given by $\Gamma_H\,=\,2.41\,\GeV,\,\Gamma_A\,=\,52.5\,\GeV,\,\Gamma_a\,=\,26.5\,\GeV,\,\Gamma_{H^\pm}\,=\,12.1\,\GeV$, rendering all width/ mass ratios $\lesssim\,6\,\%$.
The production cross sections at 3 \TeV are given by $\sigma_{HA}\,=\,0.512\,\fb,\,\sigma_{Ha}\,=\,0.390\,\fb$. The $H$ decays to $t\,\bar{t}$ with a branching ratio of $93\%$, while $\chi\bar{\chi}$ final states have branching ratios of $64\,\%/ 95\,\%$ for $A/a$, respectively. This leads to an overall estimated production cross section of around $0.65\,\fb$, using factorization and neglecting possible interference effects. 
The next step in this investigation would be a detailed simulation of signal and background in a CLIC-like environment, as e.g. performed in \cite{Kalinowski:2018kdn,deBlas:2018mhx} for the Inert Doublet Model. This is in the line of future work.
\section{Summary and outlook}\label{sec:summ}
\vspace{-1mm}
In this work, we for the first time presented a scan for the THDMa that lets all 12 free parameters of that model float freely, within ranges that were chosen to optimize scan performance. We have identified regions in parameter space that survive all current theoretical and experimental constraints, and provided a first estimate of possible production cross sections within this model at future $e^+ e^-$ facilities, with a focus on signatures not present in a standard THDM. We have included direct LHC search results using a simplified approach, with maximal cross sections as an upper limit. Such searches typically constrain the low-mass regions of parameter space, where the dark sector masses are $\lesssim\,500\,\GeV$, as well as regions with mixings $|\sin\theta|\,\gtrsim\,0.15$. In general, however, clear boundaries in two-dimensional planes are not easy to identify. An exception are constraints from B-physics observables, that expecially impose a lower limit on the charged mass $\,\sim\,800\,\GeV$ that varies with $\tan\be$. Furthermore, constraints from electroweak precision observables in the form of oblique parameters pose relatively strong constraints on the mass differences in the THDMa scalar sector for novel scalars, leading to a general lower mass scale $\sim\,500\,\GeV$. Requring the relic density to lie below the current experimental measurement furthermore poses strong constraints on $|m_a-2\,m_\chi|$. \\
We consider this work as a first step towards a more detailed analysis of this model. This includes more detailed recasts of the current LHC experimental results, including possible forecasts for HL-LHC sensitivity, as well as more detailed investigations at future lepton machines, as e.g. CLIC or possible muon-colliders. First estimates for the former show that the most promising novel channel could be $t\,\bar{t}+\slashed{E}$ in a CLIC-like scenario with a 3 \TeV~center-of-mass energy.

Finally, in view of recent events \cite{Abi:2021gix}, one might wonder whether the anomalous magnetic momentum of the muon can be explained by the model discussed here. Contributions should be similar to the ones of a general THDM of type II, as e.g. discussed in \cite{Cherchiglia:2017uwv,Ferreira:2021gke}, with additional contributions of the second pseudoscalar $a$, including appropriate rescaling. However, from the above work it becomes obvious that pseudoscalar masses would have to be light, $\lesssim\,200\,\GeV$, with additionally relatively large $\tan\be$ values $\gtrsim\,15$, in order to account for the observed experimental value. Such points have not been included in our analysis here. A quick check using the next-to-leading order approximation given in \cite{Cherchiglia:2016eui} shows that maximal values of $\Delta a_\mu^\text{BSM}\,\sim\,2.5\,\times\,10^{-12}$ can be achieved with e.g. the points which pass all other (including experimental LHC) constraints, which is around three orders of magnitude lower than the observed discrepancy. In \cite{Haller:2018nnx}, regions are shown in the $\lb m_{H^\pm},\tan\be\rb$ plane; here, for charged masses $\gtrsim\,800\,\GeV$, $\tan\be\,\gtrsim\,7$ seems to be required for $68\%$ confidence level agreement with the result from \cite{Bennett:2006fi}. This could a priori be an interesting region to investigate, which we will leave for future work.
\noindent
\section*{Acknowledgements}
The author wants to sincerely thank J. Kalinowski, W. Kotlarski, D. Sokolowska, and A.F. Zarnecki for useful discussions in the beginning of this project, and especially W. Kotlarski for help with the setup of the Sarah/ Spheno interface. Further thanks go to M. Misiak and U. Nierste for discussions regarding bounds from B-physics observables, and M. Goodsell as well as the authors of \cite{Abe:2019wjw} for advice. This research was supported in parts by the National Science Centre, Poland, the HARMONIA project under contract UMO-2015/18/M/ST2/00518 (2016-2021), and the OPUS project under contract UMO-2017/25/B/ST2/00496 (2018-2021).
\begin{appendix}
\section{Potential parameters}\label{app:invert}
The potential parameters can be expressed in terms of the free input parameters as follows:
\begin{eqnarray}
\mu_3 &=& \frac{1}{2} s_{2\alpha} (m_h^2 - m_H^2) - m_{H^\pm}^2 s_{2\beta}+\frac{v^2}{2} \lambda_3 s_{2\beta}\\
\lambda_1 &=& \frac{s_\alpha s_{\alpha+\beta}}{2v^2 c_\beta^3} m_h^2 + \frac{c_\alpha c_{\alpha+\beta}}{2v^2 c_\beta^3}m_H^2 - \frac{t_\beta^2}{v^2} m_{H^\pm}^2+\frac{t_\beta^2}{2} \lambda_3\\
\lambda_2 &=& \frac{c_\alpha s_{\alpha+\beta}}{2v^2 s_\beta^3} m_h^2 - \frac{s_\alpha c_{\alpha+\beta}}{2v^2 s_\beta^3}m_H^2 - \frac{1}{v^2 t_\beta^2} m_{H^\pm}^2+\frac{1}{2t_\beta^2} \lambda_3\\
\lambda_5 &=& -\frac{1}{2v^2} \left( -2 m_{H^\pm}^2 + v^2 \lambda_3 + m_A^2 c_\theta^2 + m_a^2 s_\theta^2 + (m_h^2-m_H^2)\frac{s_{2\alpha}}{s_{2\beta}} \right)\\
\lambda_4 &=& \frac{1}{v^2} \left(  m_A^2 c_\theta^2 +  m_a^2 s_\theta^2 - v^2 \lambda_3 +  (m_H^2-m_h^2)\frac{s_{2\alpha}}{s_{2\beta}} \right)\\
{b_p} &=& \frac{(m_a^2-m_A^2) c_\theta s_\theta}{v} \\
m_P^2 &=& m_a^2 c^2_\theta +m_A^2 s^2_\theta - v^2(\lambda_{p1} c_\beta^2  + \lambda_{p2} s_\beta^2)
\end{eqnarray}
\end{appendix}

\end{document}